\documentclass[usenatbib]{mnras}
\usepackage{natbib}
\usepackage{aas_macros}
\usepackage{graphicx}
\usepackage{lipsum}
\usepackage{xcolor}
\usepackage{amssymb}
\usepackage[titletoc]{appendix}
\usepackage{longtable}
\usepackage{lscape}
\usepackage{subfig}

\newcommand{\Chandra}{\textit{Chandra}}
\newcommand{\NuSTAR}{\textit{NuSTAR}}
\newcommand{\XMM}{\textit{XMM-Newton}}

\title[Cluster Temperatures from \NuSTAR\ and \Chandra]{A Systematic Comparison of Galaxy Cluster Temperatures Measured with \NuSTAR\ and \Chandra}
\author[A. N. Wallbank et al.]
{\parbox{\textwidth}{A. N. Wallbank$^1$, B. J. Maughan$^1$\thanks{E-mail: ben.maughan@bristol.ac.uk}, F. Gastaldello$^2$, C. Potter$^3$, and D. R. Wik$^3$} \vspace{0.4cm}\\
\parbox{\textwidth}{$^1$H. H. Wills Physics Laboratory, University of Bristol, Tyndall Ave, Bristol BS8 1TL, UK.\\
  $^2$INAF – IASF Milano, Via E. Bassini 15, 20133 Milano, Italy \\
  $^3$Department of Physics \& Astronomy, University of Utah, 115 South 1400 East, Salt Lake City, UT 84112, USA}}

\date{Accepted XXX. Received YYY; in original form ZZZ}

\pubyear{????}

\begin{document}
\label{firstpage}
\pagerange{\pageref{firstpage}--\pageref{lastpage}}
\maketitle

\begin{abstract}
Temperature measurements of galaxy clusters are used to determine their masses, which in turn are used to determine cosmological parameters. However, systematic differences between the temperatures measured by different telescopes imply a significant source of systematic uncertainty on such mass estimates. We perform the first systematic comparison between cluster temperatures measured with \Chandra\ and \NuSTAR. This provides a useful contribution to the effort of cross-calibrating cluster temperatures due to the harder response of \NuSTAR\ compared with most other observatories. We measure average temperatures for 8 clusters observed with \NuSTAR\ and \Chandra. We fit the \NuSTAR\ spectra in a hard ($3-10$ keV) energy band, and the \Chandra\ spectra in both the hard and a broad ($0.6-9$ keV) band. We fit a power-law cross-calibration model to the resulting temperatures. At a \Chandra\ temperature of 10 keV, the average \NuSTAR\ temperature was $(10.5\pm3.7)\%$ and $(15.7\pm4.6)\%$ lower than \Chandra\ for the broad and hard band fits respectively. We explored the impact of systematics from background modelling and multiphase temperature structure of the clusters, and found that these did not affect our results. Our sample are primarily merging clusters with complex thermal structures so are not ideal calibration targets. However, given the harder response of \NuSTAR\ it would be expected to measure a higher average temperature than \Chandra\ for a non-isothermal cluster, so we interpret our measurement as a lower limit on the difference in temperatures between \NuSTAR\ and \Chandra.
\end{abstract}


\section{Introduction}
Galaxy clusters are the largest gravitationally bound objects in the Universe. Massive galaxy clusters are ideal targets for studying cosmology as their number density is sensitive to cosmological parameters \citep[see e.g.][for a review]{all11}. Cluster masses are essential for such cosmological studies. However, since clusters are dominated by dark matter, their masses must be inferred from observable properties \citep[the main methods used are reviewed in][]{pratt19}. The X-ray emission from the intra-cluster medium (ICM) provides several observable quantities which may be used for this purpose. These include temperatures and densities of the ICM, radial profiles of which can be used to infer the total mass of a cluster assuming hydrostatic equilibrium. Where the X-ray data quality does not allow detailed measurements, scaling relations between ICM quantities (such as luminosity, temperature or gas mass) may be used to estimate masses if they are accurately calibrated.

There has been significant attention paid recently to the accuracy of cluster masses obtained from hydrostatic equilibrium analyses. If hydrostatic masses were systemtatically underestimated, this could help to resolve the discrepancy that has been observed in some  of the cosmological parameters (principally $\sigma_8$) obtained from \textit{Planck} measurements of the cosmic microwave background (CMB) anisotropies versus those obtained from number densities of massive clusters detected by \textit{Planck} \citep{plan18}.

Hydrostatic masses may be underestimated if there is significant non-thermal pressure support in the ICM, since the X-ray observations only measure the thermal pressure \citep{lag10,barn21,ett22}. This is known as the hydrostatic bias, and conventionally refers to the amount by which hydrostatic masses are biased low compared to the true mass. Comparisons between hydrostatic masses and masses obtained via weak gravitational lensing measurements suggest that that a hydrostatic bias may be present but the size of the bias remains uncertain. Most analyses point towards hydrostatic masses being underestimated by $\sim(0-20)\%$ \citep{von14,oka16,her20,sal21,log22}, but underestimates as large as $(30-40)\%$ may be required to fully align the \textit{Planck} CMB and cluster number counts constraints on $\sigma_8$ \citep{planck20}.

The calibration of X-ray telescopes is another possible source of
systematic uncertainty on hydrostatic masses. In particular if ICM
temperatures were systematically underestimated, this would lead to
underestimates of hydrostatic masses. Absolute calibration of X-ray
telescopes in orbit is challenging due to the lack of standard,
unvarying astrophysical calibration sources, and the cross-calibration
of different X-ray telescopes reveals some inconsistencies. One of the
key instrumental properties that needs to be calibrated is the
effective area of the X-ray telescopes. This is energy-dependent, and
miscalibration at different energies would distort the slope of
observed X-ray spectra, potentially biasing quantities such as ICM
temperatures that are derived from spectral fits. Restricting spectral
fitting to different X-ray energy bands can then help to reveal any
energy-dependent problems with the effective area calibration. As
discussed below, it is difficult to ascribe temperature differences
directly to the effective area calibration due to the presence of
other instrumental effects, such as the point spread function (PSF) in
the case of extended sources, and detector gain. The contributions of
multiple temperature components in the ICM to the observed spectrum
also complicate the interpretation of temperature differences.

Cross-calibration work has been supervised by the International Astrophysical Consortium for High Energy Calibration (IACHEC), including the use of galaxy clusters as calibration sources. \citet{nev10} and \citet{sch15} both performed a cross-calibration using galaxy clusters observed with both \textit{Chandra} and \textit{XMM-Newton}, focussing on the ICM temperature measurements. \citet{nev10} found that temperatures measured with \textit{XMM-Newton} were systematically lower than those measured with \textit{Chandra} by 10-15\% on average, but the difference was not significant if the spectral fitting was restricted to higher X-ray energies ($\gtrsim 2$ keV).

\citet{sch15} found qualitatively similar results using a much
  larger sample, with the disagreement between \textit{Chandra} and
\textit{XMM-Newton} increasing with cluster temperature.
\textit{Chandra} temperatures were $\sim20\%$ higher than those
measured using \textit{XMM-Newton} at a cluster temperature of $10$
keV. Again the discrepancy was found to be dominated by the inclusion
of softer X-rays ($\lesssim 2$ keV) in the spectral fitting. This work
also tested the internal cross-calibration of the observatories'
instruments. There was good agreement between temperatures measured by
\Chandra\ Advanced CCD Imaging Spectrometer (ACIS) I and S arrays
(which use the same optics) but inconsistencies between temperatures
from the three \XMM\ European Photon Imaging Camera (EPIC) detectors
(which each use a different telescope).

A cross-calibration of \textit{Suzaku} and \textit{XMM-Newton} using galaxy clusters has also been performed \citep{ket13}. The results showed that \textit{Suzaku} systematically measured the temperature of galaxy clusters to be 2-6\% lower than those from \textit{XMM-Newton} when the analysis was performed in the hard energy band of 2.0-7.0 keV and up to a 12\% difference for temperatures measured in the soft energy band of 0.5-2.0 keV. However, it was determined that up to half of this difference could be caused by \textit{Suzaku's} PSF scattering (typically softer) photons from the core regions.

The results from these cross-calibrations highlight the uncertainty in measurements between the different X-ray observatories and the importance of understanding how different detectors are calibrated.

\NuSTAR\ presents a relatively untapped opportunity for the cross-calibration of cluster temperatures. The hard response of \textit{NuSTAR} makes it more sensitive to the exponential cut-off of the bremsstrahlung continuum that dominates the ICM emission for hot clusters. This enables it to more accurately measure temperatures of hot clusters compared to e.g. \textit{Chandra}, \textit{XMM-Newton} or \textit{Suzaku}. \NuSTAR\ has been used for spectral analysis of a relatively small number of galaxy clusters, often with the aim of constraining any non-thermal component of the emission \citep[e.g.][]{wik14, xmm-nus19,roj20}.

In this paper, we present the first systematic comparison of galaxy cluster temperatures measured with \textit{NuSTAR} and \textit{Chandra}. We compare the temperatures of a sample of eight galaxy clusters which have been observed with both \textit{NuSTAR} and \textit{Chandra}.
The paper is organised as follows. We give an overview of the cluster sample in section \ref{sec:sample}. Section \ref{sec:data_anly} discusses \NuSTAR\ and \Chandra\ data processing as well as the background and spectral analysis. The temperature measurements and cross-calibration results are presented in section \ref{sec:results}. The results are then discussed in more detail in section \ref{sec:disc}, and our conclusions are presented in section \ref{sec:summary}.
Throughout this paper, when referring to \Chandra\ and \XMM\ temperatures, we mean those measured with the \Chandra\ ACIS and \XMM\ EPIC detectors.

\section{Cluster sample}
\label{sec:sample}
In principle, galaxy clusters make good X-ray calibration sources since they are bright and non-varying, and generally unaffected by pile-up. However, the ICM can be thermally complex with multiphase gas projected along the line of sight. If a model with a single temperature component is fitted to an X-ray spectrum from a region of ICM, then the measured temperature will be an average of the different temperature components of the ICM along the line of sight, weighted by the emissivity of each component combined with the energy-dependent effective area of the telescope \citep{maz04, viklinin06}. If a galaxy cluster were perfectly isothermal, then all correctly-calibrated X-ray telescopes would recover the same temperature. However, when the observed spectrum comprises a combination of temperature components, then even perfectly calibrated telescopes would recover different average temperatures if their sensitivities were different at different X-ray energies. For example, a telescope with a hard response (like \textit{NuSTAR}) would recover a higher average temperature than a telescope with a softer response (like \textit{Chandra}).

For cross-calibration purposes it would be ideal, therefore, to select a sample of relaxed galaxy clusters which contain large volumes of approximately isothermal gas. However, the relatively small number of clusters that have been observed by \textit{NuSTAR} are almost exclusively merging clusters. This is by design, since such systems are expected to contain very hot ICM regions related to merger shocks and radio halos (with possible associated non-thermal X-ray emission) so make good targets for \textit{NuSTAR} observations.

As of January 2020, the \textit{NuSTAR} public archive contained eight galaxy clusters that we deemed were suitable for our purposes. In particular, we required the clusters to be bright but sufficiently distant that the majority of the ICM emission was contained within the field of the Focal Plane Modules, giving a sample of eight clusters. All of these clusters had previously been observed by \textit{Chandra}, many of them with a large number of observations. A subset of the available observations were chosen to provide a similar data quality to the \NuSTAR\ data (as will be seen later, the statistical uncertainty on the \Chandra\ temperature measurements is sufficiently small that there is no benefit to utilising all of the available data). The resulting sample is presented in Table \ref{tab:clusters}.

\begin{table*}
\centering
    \begin{tabular}{lcccccccc}
    \hline
    Cluster Name & \NuSTAR\ & $t_N$ & \Chandra\ & $t_C$ & Redshift & N$_H$ & RA & Dec \\[0.5ex]
     & ObsID & ks & ObsID & ks & z & (10$^{22}$ cm$^{-2}$) & degrees & degrees  \\
    \hline\hline
    Abell 2146 & 70401001002 & 226 & 12247, 12245 & 114 & 0.232 & 0.0337 & 239.042 & 66.355  \\
    Abell 2163 & 70101002002 & 99 & 1653 & 71 & 0.203 & 0.2060 & 243.942 & -6.143\\
    Abell 2256 & 70001053002 & 86 & 16514, 16129 & 89 & 0.058 & 0.0494 & 255.975 & 78.636\\
    Abell 523  & 70301001002, 70102001004 & 184 & 15321 & 30 & 0.104 & 0.1590 & 74.779 & 8.753  \\
    Abell 665  & 70201002002, 70201003002 & 164 & 13201, 12286 & 89 & 0.183 & 0.0507 & 127.729 & 65.853  \\
    Abell 754 & 70201001002 & 104 & 10743 & 94 & 0.054 & 0.0579 & 137.346 & -9.684 \\
    1E 0657-56  & 70001055002, 70001055002 & 251 & 3184, 4986 & 121 & 0.296 & 0.0643 & 104.621 & -55.944  \\
    RX J1347.5-1145 & 70301001002, 70301001004 & 142 & 3592 & 57 & 0.451 & 0.0581 & 206.879 & -11.753  \\[1ex]
    \hline
    \end{tabular}
    \caption{Summary of the clusters analysed in this work. The columns give the cluster name, \NuSTAR\ ObsID(s) and total cleaned exposure time ($t_N$), \Chandra\ ObsID(s)and total cleaned exposure time ($t_C$), redshift, hydrogen column density \citep[taken from][]{nh13}, right ascension and declination. The coordinates given correspond to the centre of the source region used to extract the spectra for our analysis.}
    \label{tab:clusters}
\end{table*}

\section{Data Processing and Analysis}
\label{sec:data_anly}
\subsection{\textit{NuSTAR} data processing}
\label{subsec:nustar_data}
The \NuSTAR\ observations of each cluster were processed with the
latest calibration as of May 2020 (CALDB version
20200429\footnote{Subsequent to the completion of our analysis,
  \NuSTAR\ CALDB 20211020 was released, which included an update to
  the effective area \citep{mad21}. The change to the effective area
  is almost constant with energy below $10$ keV, so is not expected to
  significantly impact cluster temperature measurments. We confirmed
  this by repeating our analysis for A2146 with the new calibration
  and found that temperature changed by $2\%$.}). For four clusters
(Abell 523, Abell 665, 1E 0657-56 and RX J1347.5-1145), there were two
\NuSTAR\ observations, and in these cases, both observations were
processed and used in the analysis. The data reduction was performed
for both focal plane modules A and B (FPMA and FPMB) for each cluster.
The standard {\tt
  nupipline}\footnote{\url{https://heasarc.gsfc.nasa.gov/docs/nustar/analysis/nustar_swguide.pdf}}
processes were run on the observations, with the additional arguments
{\tt saamode=STRICT} and {\tt TENTACLE=yes}, closely following the
analysis done in \citet{xmm-nus19}.

\textit{NuSTAR} has a complex background, which must be accounted for when performing spectral analyses. The approach we use closely follows that described in \citet{wik14} and consists of extracting background spectra from the target observation, fitting a detailed model to those, and then simulating background spectra for the source region based on this model.

We briefly summarise the main components of the \textit{NuSTAR} background below. A  more detailed description can be found in \citet{wik14, Har13, mads17}.

The focused cosmic X-ray background (``fCXB") is due to unresolved sources in the field of view (FOV) and can be important below $\sim$15 keV. However, $\sim90\%$ of the cosmic X-ray background (CXB) photons detected in the FOV are due to stray light that seeps through the aperture stops, creating a spatial gradient across the FOV. This is the dominant background component at energies below $\sim 15$ keV, and is known as the ``aperture" background. Stray light (from the CXB, the Sun and Earth's albedo) is also present in the FOV due to reflection from surfaces of the spacecraft. This component can undergo significant fluctuations due to solar activity at the observation time.
Above 15-20 keV, the instrument or internal background dominates. This component differs between the FPMA and FPMB detectors but is spatially uniform and consists of a continuum plus strong fluorescent and activation lines.

To model the background {\tt nuskybgd} was used \citep{wik14}. This tool fits a model to background spectra extracted from the target observation and then produces a simulated background at the detector position of the source. This includes the spatial variation of the different background components where necessary, and geometric corrections for the different region sizes. This simulated background is then used in the analysis of the source spectrum.

\begin{figure}
    \includegraphics[width=\columnwidth]{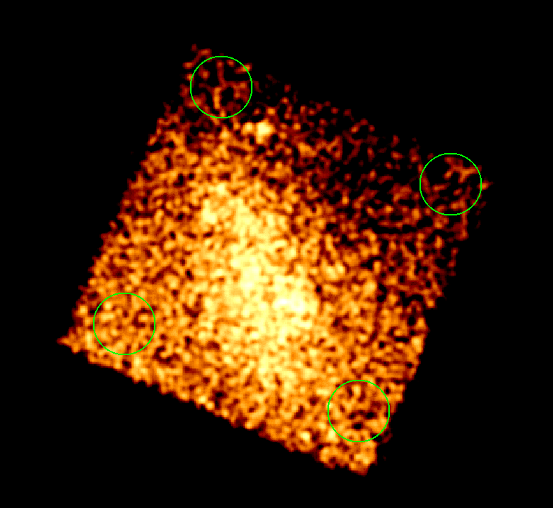}
      \caption{\textit{NuSTAR} image of Abell 523, ObsID 70102001002, showing the four regions used to model the background with {\tt nuskybgd}. The image is from detector FPMA in the energy band 3-20 keV and has been smoothed with a Gaussian with $\sigma=6.25"$. The background has not been subtracted in this image, and the strong gradient in the aperture background is visible across the field.}
    \label{fig:A523_2_bgd_reg}
\end{figure}

Background regions were defined in each corner of the FOV, avoiding areas of high cluster emission when possible, similar to the approach used in \citet{xmm-nus19}. For observations where the cluster was located in one of the corners of the FOV, background regions were defined in the remaining three corners. The same regions were used for both FPMA and FPMB. Figure \ref{fig:A523_2_bgd_reg} presents an example of an \textit{NuSTAR} image of Abell 523, in the energy band of $3-20$ keV, indicating the regions used to extract background spectra.

Because the target clusters are bright and relatively nearby, their emission fills a large fraction of the \NuSTAR\ FOV, and so it is possible that emission from the cluster is present in the background regions at a non-negligible level. We accounted for this by including an additional APEC \citep{smi01} component in the background model to describe the ICM emission. The temperature, abundance and normalisation of this component were free to fit (with the values tied for matching regions on the FPMA and FPMB detectors). In general, the parameters of this model were not well constrained, and the normalisation was degenerate with other background model components.
(The process was also performed with the temperature and abundance of the APEC component in the background region fixed at the values measured for the cluster, with consistent results.)

In all cases, the impact of this additional APEC component in the
background model on the final temperatures measured for the clusters
was negligible (the temperatures changed by $<3\%$ compared to the
case where no additional APEC component was included, and the change
was random in direction for the eight clusters rather than a
systematic increase or decrease). Our final temperature measurements
used the background models where the additional APEC component was
included, but this has no significant effect on our results.

Once the simulated background spectra were produced, the source spectra were then extracted and associated response matrices (RMF) and auxiliary response files (ARF) were produced using {\tt nuproducts}. The parameter {\tt extended=yes} was used, which weights the ARF based on the distribution of events within the chosen extraction region\footnote{This weighting was calculated using events detected over the whole 3-79 keV energy band (i.e. the keywords {\tt pilowarf} and {\tt pihigharf} were not used), which results in the distribution being flatter than the cluster emission. This could have the effect of weighting the ARF too much towards the outer parts of the field of view and hence underestimating the effective area at higher energies. However the effect is negligible for our analysis; the effective area at 10 keV changes by $\sim 0.1\%$ when the ARF is weighted by the events in the $3-10$ keV band.}. The cluster source spectra were extracted from a circular region of radius $250"$ centred on the emission centroid determined from the \NuSTAR\ image. This radius was chosen to produce a high signal to noise spectrum for the clusters studied, and was used for consistency for all clusters (although, as seen in Appendix \ref{appen:A}, the results are not sensitive to the choice of region). The same regions were used for both FPMA and FPMB detectors.

\subsection{\textit{Chandra} data processing}
\label{subsec:chan}
The \textit{Chandra} data were processed following the reduction and analysis procedures presented in \citet{mau08, mau12}, and we summarise the main steps below. For all clusters, \Chandra\ ACIS-I observations were used. In this work \textit{CIAO} version 4.12 and \textit{CALDB} version 4.9.1 was used.

The data were reprocessed from level 1 events and lightcurves were produced (in the $0.3-12$ keV band) and filtered to remove periods of high background. Blank-sky backgrounds events lists were used in the spectral analysis and were processed in a consistent way to the source observations. Exposure corrected images in the $0.3-7$ keV energy band were used for the detection of point-like sources. All the point sources were excluded from the subsequent analysis.

We note that in all cases, any point sources detected in the
\textit{Chandra} data that fell within the $250"$ source region used
for the spectral analysis were not resolved in the \textit{NuSTAR}
observations, and so were not excluded in the \textit{NuSTAR} spectral
analysis. However, in all cases these point sources were faint. Within
the $250"$ source spectrum region, the count rate from resolved point
sources measured with \textit{Chandra} was at most $3\%$ of the total
emission, in any of the energy bands used for the \Chandra\ and
\NuSTAR\ spectral analysis. We verified that excluding the point
sources in the \Chandra\ analysis did not impact the temperatures by
measuring the temperature of A2146 (which had the maximum $3\%$ point
source contamination) without excluding the sources. Retaining the
point sources increased the \Chandra\ temperature by $1\%$.

Radial surface-brightness profiles of the cluster emission were used to define the radii beyond which the cluster emission was negligible compared to the background in the \Chandra\ observations. This enabled the definition of local background regions within which the local and blank-sky background spectra could be compared.

The background spectra were normalised to match the count rate in the hard X-ray band ($9.5-12$ keV). The residuals between the local and blank-sky background spectra were then primarily due to differences in soft Galactic foreground emission at $<2$ keV. To account for this, the residuals were fit with an unabsorbed APEC model with a temperature of $0.18$ keV and solar abundance, which was then included in the subsequent analysis of the source spectrum.

Since the target clusters are bright and relatively nearby, in some cases the cluster emission fills a large fraction of the \Chandra\ FOV, and may be present in the regions defined for the background analysis. As discussed in \textsection \ref{sec:disc}, due to the high signal-to-noise of the source spectra, any cluster emission in the background region is not expected to significantly influence the temperature measurement.

\begin{figure}
    \subfloat{\includegraphics[width=\columnwidth]{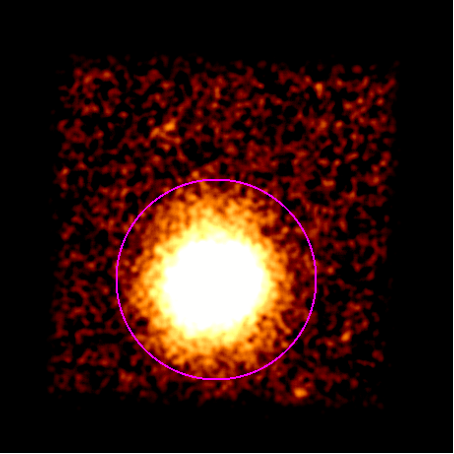}} \\
    \subfloat{\includegraphics[width=\columnwidth]{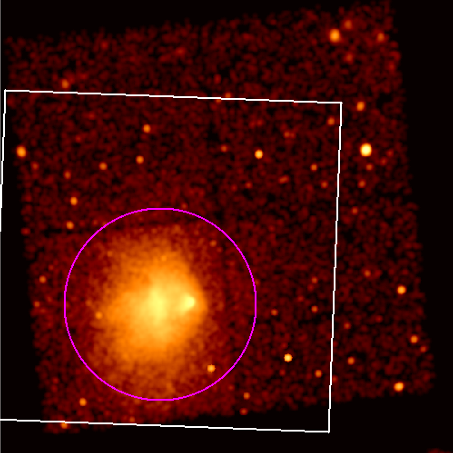}}
    \caption{\NuSTAR\ (top) and \Chandra\ (bottom) image of 1E 0657-56, ObsID 70001055002 and 3184, respectively. The \NuSTAR\ observation is a {\tt nuskybgd} background-subtracted image in the $3-20$ keV energy band, while the \Chandra\ image is in the $0.7-2.0$ keV energy band. Both images are smoothed with a Gaussian with $\sigma=6.25"$. The magenta circle is the region in which the source spectra were extracted. The white square shows the region covered by the \NuSTAR\ detectors overlaid on the \Chandra\ image.}
    \label{fig:bul_ob}
\end{figure}

The cluster spectra were then extracted for each observation using the $250"$ circular source regions defined for the \NuSTAR\ analysis. For illustration, a comparison of the \NuSTAR\ and \Chandra\ images of 1E 0657-56 is shown in Figure \ref{fig:bul_ob}, with the spectral extraction region shown in magenta, and the \NuSTAR\ FOV overlaid on the \Chandra\ image in white. The images of all clusters are presented in Appendix \ref{appen:B}.

\subsection{Spectral fitting}
\label{subsec:spec_fit}
The cluster spectra were modelled with a single temperature, absorbed APEC model convolved with the appropriate ARF and RMF for the regions used. The spectra were fit in XSPEC \citep{ard96}, using the {\tt cstat} statistic, with the normalisation, temperature and metallicity free to vary when fitting, while the N$_H$ and redshift were frozen. The absorption was modelled with \textit{phabs}, and the values for N$_H$ were derived from \citet{nh13}, with XSPEC's default abundance table from \citet{and89}.

For the \NuSTAR\ analysis, the spectra were fit between $3-10$ keV (hereafter referred to as the hard band), which is close to optimal given the instrument response and typical cluster spectrum. For \Chandra, the spectra were fit in the hard band for consistency with \NuSTAR, and in the $0.6-9$ keV band (hereafter referred to as the broad band), which is more suitable for \Chandra\ analyses given the softer response.

For the \textit{NuSTAR} observations, the temperature was measured for the FPMA and FPMB detectors individually, as well as by fitting a single source model to the spectra from the two detectors simultaneously. Clusters with two \textit{NuSTAR} observations (Abell 523, Abell 665, 1E 0657-56 and RX J1347.5-1145) were fit individually and combined, with the same being done for \textit{Chandra} data when two observations were used (Abell 665, Abell 2146, Abell 2256, and 1E 0657-56).

\subsection{Notes on individual clusters}
\label{subsec:clusters}
In this section we note any departures that were required from the standard analysis described in the preceding sections.
\subsubsection{Abell 754}
For Abell 754, the location of the cluster within the \NuSTAR\ FOV meant that the $250"$ circular source region extended off the edge of the detectors in both observations. The part of the region that fell off the detector in any of the data was excluded from all of the \NuSTAR\ and \Chandra\ spectral extractions.

\subsubsection{Abell 2163}
\label{subsubsec:A2163}
For the \textit{Chandra} analysis of A2163, the lightcurve showed a gradient over the duration of the observation, indicating possible a low-level background flare. The lightcurve was manually cleaned to select a 20 ks period where the background appeared stable. After this filtering, the comparison between the local and blank-sky background showed a strong excess in the local background below 2 keV, with some hints of excess emission up to $\sim 7$ keV, which could indicate residual flare contamination. Abell 2163 is also located near the North Polar Spur (NPS), giving it a particularly complex soft background component and strong N$_H$ gradient across the cluster \citep{pratt01,bour11,roj20}.

As a result we consider the \Chandra\ temperature measurement for this cluster to be less reliable than for the other clusters in our sample, particularly for the spectral fit in the broad band, and hence we exclude the broad-band temperature from our main analysis.

\subsubsection{Abell 2256}
\label{subsubsec:A2256}
For this cluster, the local background spectrum in the \Chandra\ data had a significant excess above the blank-sky background below $\sim$2 keV. Unlike Abell 2163 there was no indication of residual flaring in the lightcurve. This is a bright, nearby cluster with significant emission from the ICM across most of the \Chandra\ FOV (see Figure \ref{fig:app-images-1}). We thus interpret the excess in the local background as being due to emission from the ICM and/or a particularly strong soft Galactic foreground, which was not modelled completely by the APEC component in the standard data processing. To account for this, the temperature of the APEC component used to model the soft Galactic foreground was unfrozen, and the best fitting values (0.86 keV and 0.94 keV for observations 16514 and 16129, respectively) were used in the subsequent analysis.

We note that if the excess in the local background were due to contamination of the background spectrum by emission from the ICM, then this could lead to the flux of the soft Galactic foreground model component being overestimated and hence the cluster temperature being overestimated. However, given the very high signal to noise of the cluster spectrum within $250"$, this is not expected to have a significant effect.

\section{Results}
\label{sec:results}
\subsection{Temperatures}

\begin{table*}
    \centering
    \large
    \begin{tabular}{l c c c}
    \hline
    Cluster Name & $kT_{C,(0.6-9)}$ & $kT_{C}$ & $kT_{N}$\\ [0.5ex]
    & keV & keV & keV \\
    \hline\hline
    Abell 2146 & $7.08\pm0.14$ & $8.90\pm0.66$ & $6.83\pm0.11$ \\
    Abell 2163 & $16.36\pm0.70^\dagger$ & $12.23\pm1.15$ & $9.72\pm0.13$ \\
    Abell 2256 & $7.81\pm0.15$  & $7.38\pm0.41$ & $6.42\pm0.08$ \\
    Abell 523  & $5.30\pm0.36$ & $7.15\pm2.55$  & $4.94\pm0.11$  \\
    Abell 665  & $8.66\pm0.23$  & $8.29\pm0.62$ & $7.45\pm0.12$  \\
    Abell 754 & $9.09\pm0.17$  & $9.25\pm0.47$ & $8.58\pm0.14$ \\
    1E 0657-56  & $13.57\pm0.36$ & $14.57\pm0.96$ & $12.58\pm0.18$ \\
    RX J1347.5-1145 & $14.71\pm0.46$  & $15.80\pm1.09$  & $12.64\pm0.28$\\ [1ex]
    \hline
    \end{tabular}
    \caption{The temperature measurements for the cluster sample. $kT_{C,(0.6-9)}$ denotes the temperatures from the \textit{Chandra} data when fit in the 0.6-9 keV energy band, while $kT_C$ denotes the \Chandra\ temperatures measured in the 3-10 keV band. $kT_{N}$ are the temperatures measured from the \textit{NuSTAR} data when fit in the 3-10 keV energy band. $^\dagger$We consider the $kT_{C,(0.6-9)}$ measurement to be unreliable for this cluster (see \textsection \ref{subsubsec:A2163})}.
    \label{tab:main_temp}
\end{table*}

The global temperatures measured for each cluster are presented in Table \ref{tab:main_temp}. $kT_{C,(0.6-9)}$ denotes the temperatures from the \textit{Chandra} data when fit in the broad band, while $kT_C$ denotes the \Chandra\ temperatures measured in the hard band. $kT_{N}$ denotes the temperatures measured from the \textit{NuSTAR} data when fit in the hard band. The temperatures reported here are from joint spectral fits to all available detectors and observations for each cluster. When fit individually, the differences in temperatures between detectors and observations for a given cluster were consistent with the statistical uncertainties.

\begin{figure}
    \includegraphics[width=\columnwidth]{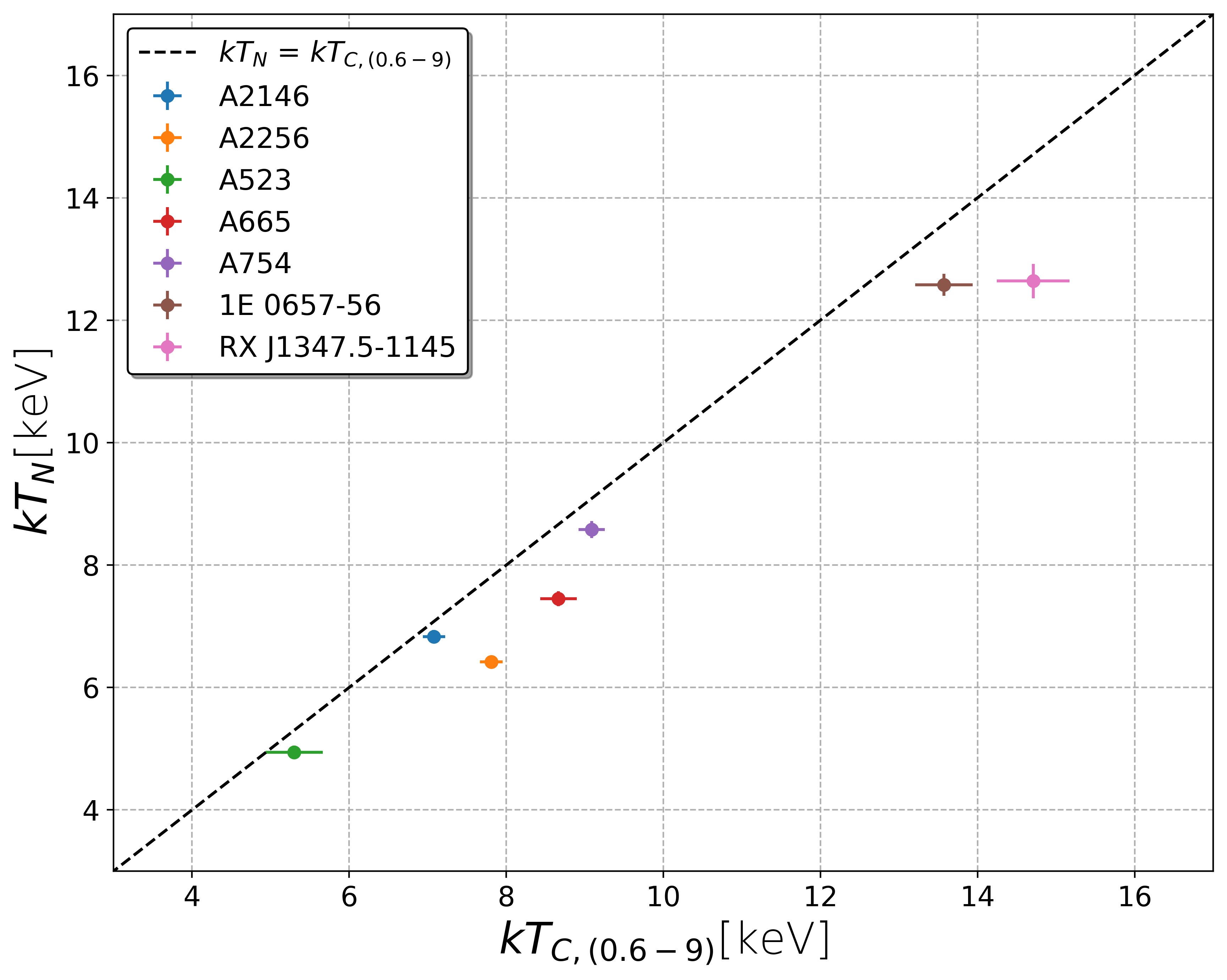}
    \caption{Cluster temperatures measured by \textit{NuSTAR} in the hard ($3-10$ keV) band are plotted against those measured by \textit{Chandra} in the broad ($0.6-9$ keV) band. Note that Abell 2163 is omitted from this plot as the \Chandra\ broad-band temperature was considered to be unreliable (see \textsection \ref{subsubsec:A2163}).}
     \label{fig:Nu_chan}
\end{figure}

Figure \ref{fig:Nu_chan} presents the comparison between the hard-band \NuSTAR\ temperatures and those measured with \Chandra\ in the broad band. This is intended to reflect the optimal energy bands used for temperature measurements with each observatory. This comparison clearly shows that the temperatures measured by \textit{Chandra} are systematically higher than those measured by \textit{NuSTAR} for the clusters in our sample.

\begin{figure}
\centering
    \includegraphics[width=\columnwidth]{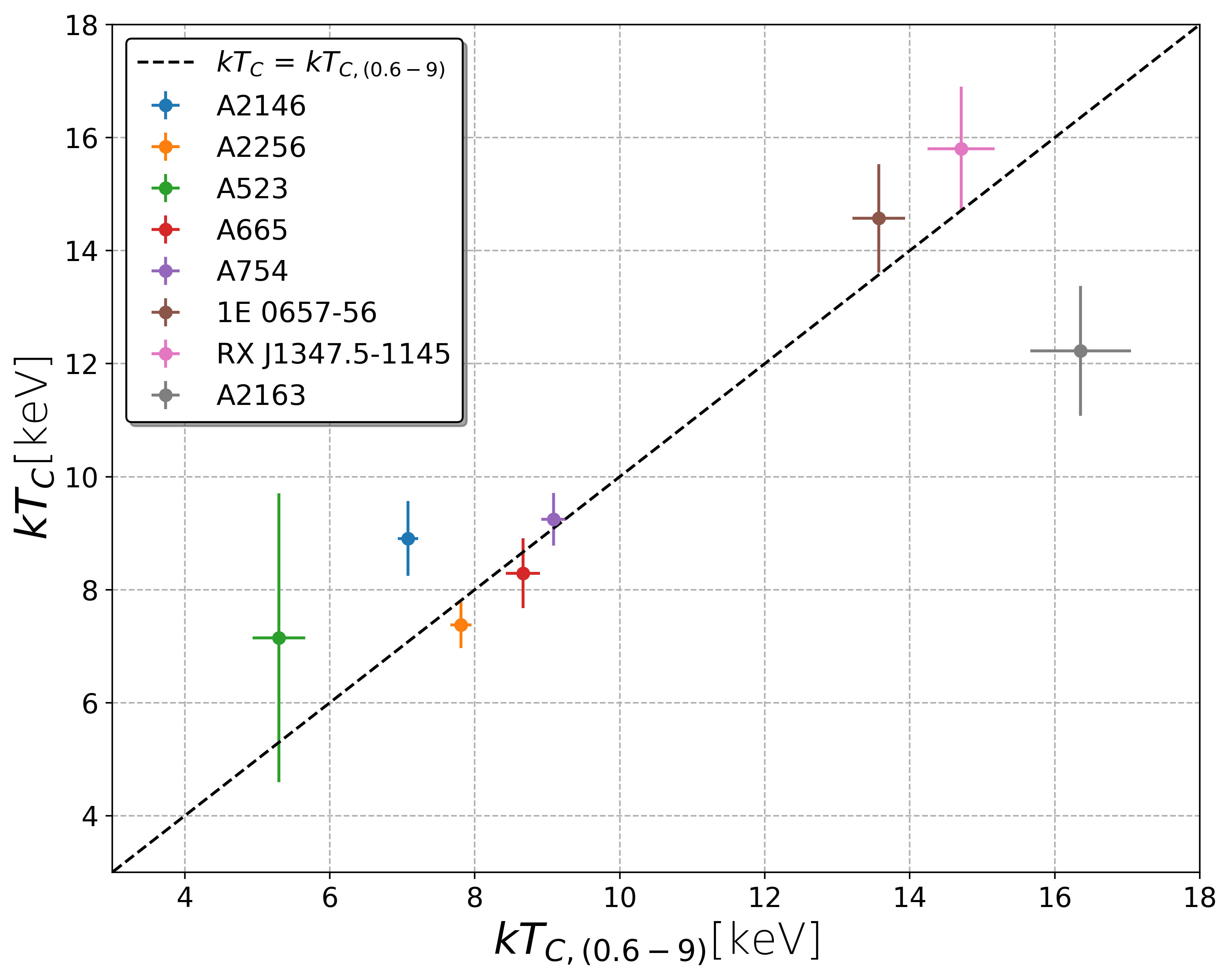}
    \caption{The \textit{Chandra} temperatures measured for spectra fit between 3-10 keV are plotted against those measured with \Chandra\ in the $0.6-9$ keV band. Note that Abell 2163 is a significant outlier due to the unreliable temperature measurement in the $0.6-9$ keV band (see \textsection \ref{subsubsec:A2163})}
    \label{fig:chan_comp}
\end{figure}

While the $0.6-9$ keV band is most appropriate for spectral analysis with \Chandra\ due to the signal-to-noise of the cluster and effective area of the telescope, a more direct comparison with \NuSTAR\ temperatures can be made using the \Chandra\ temperatures made in the same hard band used for the \NuSTAR\ analysis. In the presence of multi-phase gas in the ICM, one would expect the temperature measured with \Chandra\ in the hard band to be higher than that measured in the broad band. This comparison is shown in Figure \ref{fig:chan_comp}, which shows a tendency for the \Chandra\ hard-band temperatures to be higher than the broad-band temperatures, but the effect is not statistically significant. As noted in \textsection \ref{subsubsec:A2163}, Abell 2163 is a significant outlier due to the unreliable \Chandra\ temperature measurement in the broad band. It is also apparent from Figure \ref{fig:chan_comp} that the statistical error on the temperature is larger when fit in the hard band, particularly for the coolest cluster (Abell 523). This is unsurprising given the lower signal to noise in the \Chandra\ data above 3 keV.

The clusters in our sample are well-known and have multiple temperature measurements in the literature from \Chandra, and in many cases from \NuSTAR. However, the complex thermal structures of these clusters (again, they were mainly selected for \NuSTAR\ observation due to their disturbed dynamical state) means that temperature measurements will be sensitive to the region used for the spectral extraction and the energy range used for spectral fitting. For this reason, there is significant variation in temperature measurements from different analyses, and we forego making any detailed comparison between our temperatures and those in the literature.

\subsection{Cross-calibration}
\label{subsec:corss-cal}
\begin{table*}
    \centering
    \large
    \renewcommand{\arraystretch}{1.25}
    \begin{tabular}{c c c c c}
    \hline
    \Chandra\ Energy Band & $\alpha$  & $\beta$ & $\epsilon$ & $\Delta_{10 keV}$ \ $(\%)$\\ [0.5ex]
    \hline\hline
    0.6-9 keV & $0.952\pm0.018$ &$0.98\pm0.13$ & $0.002^{+0.003}_{-0.001}$ & $10.5\pm3.7$ \\
    3-10 keV & $0.926\pm0.024$ & $0.99\pm0.19$ & $0.002^{+0.006}_{-0.001}$ & $15.7\pm4.6$ \\ [1ex]
    \hline
    \end{tabular}
    \caption{\label{tab:fit} The cross-calibration between \textit{Chandra} and \textit{NuSTAR}. $\alpha$ and $\beta$ are the intercept and gradient from equation 1. $\epsilon$ is the intrinsic scatter from \textit{Linmix}, and $\Delta_{10 keV}$ is the percentage difference between the two detectors when the instrument X is at 10 keV. Energy Band$_C$ is the energy band in which a model was fit to the \textit{Chandra} observation spectra.}
\end{table*}

To quantify the cross-calibration between \textit{NuSTAR} and \textit{Chandra} temperatures, we follow \citet{sch15} and fit a power law model to the data. The fitting was performed using a linear model in log space using \textit{Linmix}, which performs a Markov chain Monte-Carlo sampling of the model likelihood, and accounts for measurement errors in both axes and intrinsic scatter \citep{kel07}.  The form of the scaling relationship used for this analysis is
\begin{equation}
    {\log_{10}\left( kT_{N} \right)= \beta \times \log_{10}\left(\frac{kT_{C}}{10~{\rm keV}} \right) + \alpha}
    \label{equ:scale}
\end{equation}
The fit is also performed with $kT_{C,(0.6-9)}$ in place of $kT_C$ when comparing \NuSTAR\ temperatures with broad-band temperatures from \Chandra. When using broad-band \Chandra\ temperatures, Abell 2163 was excluded as discussed in \textsection \ref{subsubsec:A2163}.

\begin{figure}
    \includegraphics[width=\columnwidth]{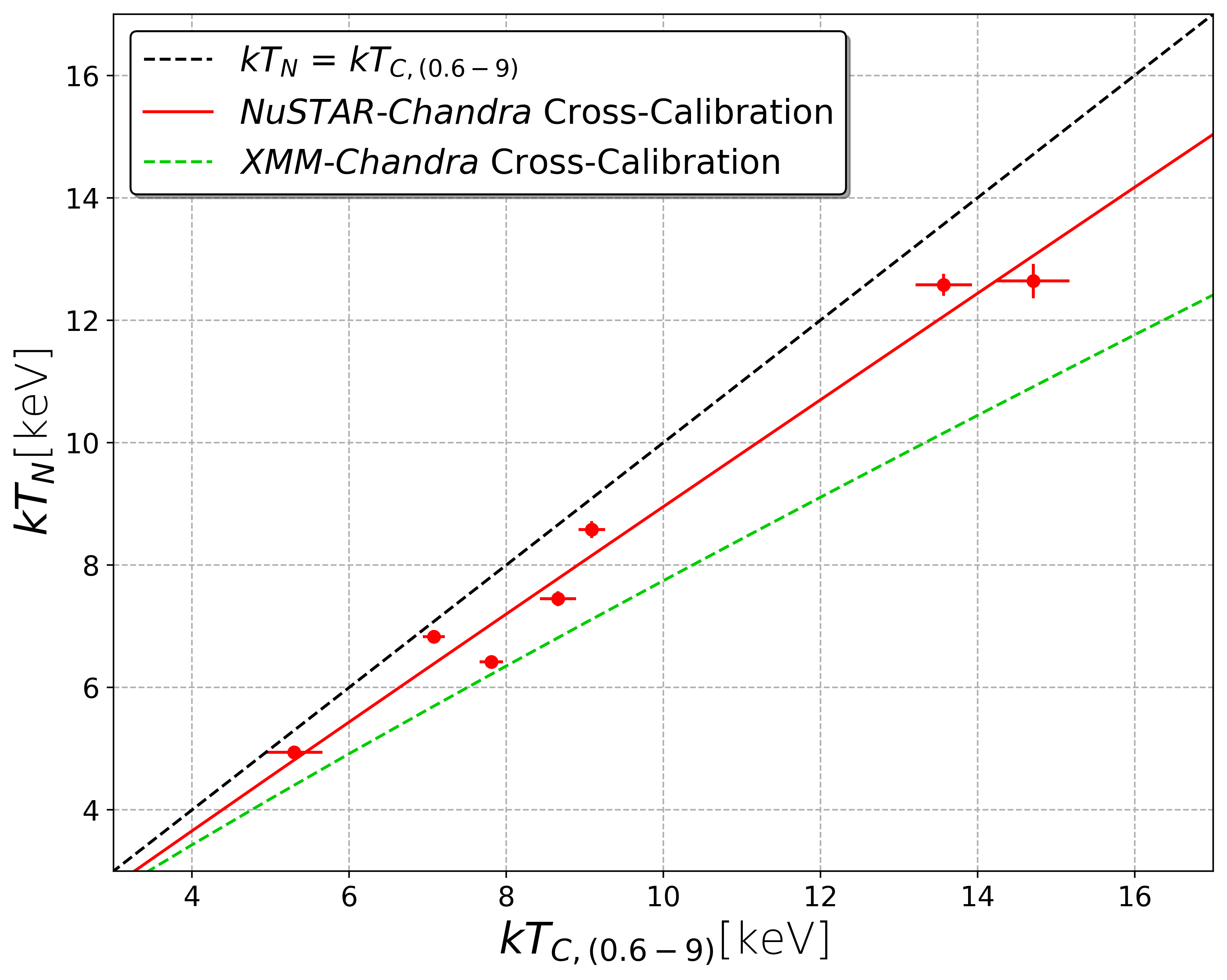}
    \caption{The cross-calibration between \textit{NuSTAR}
      temperatures measured in the $3-10$ keV band and broad-band
      \textit{Chandra} temperatures (measured in the $0.6-9$ keV
      band). The best fitting power-law model to the data is shown in
      red, and the black dashed line indicates a perfect agreement
      between the temperatures. Abell 2163 is omitted from this plot
      as discussed in \textsection \ref{subsubsec:A2163}. The green
      dashed line shows, for illustrative purposes, the best fitting
      power-law to the cross-calibration of temperatures measured with
      \textit{XMM-Newton} versus \textit{Chandra} derived in
      \citet{sch15} (we show their ``ACIS-Combined XMM Full'' model,
      measured in the $0.7-7$ keV energy band, with
      \textit{XMM-Newton} temperatures in place of \NuSTAR\
      temperatures on the vertical axis). Note that the comparison
      with the result from \citet{sch15} is not precise due to the
      different energy bands used for their temperature measurements,
      but illustrates the relative cross-calibration of the
      temperatures between the three telescopes.}
    \label{fig:log10_fit_v_XMM}
\end{figure}

The results of the fits are presented in Table \ref{tab:fit}, and the best fitting model for the broad-band \Chandra\ temperatures is plotted along with the data in Figure \ref{fig:log10_fit_v_XMM}. The uncertainties on the fit parameters are derived from the median and 16th and 84th percentiles of parameter chains sampled by \textit{Linmix}, corresponding to 1$\sigma$ uncertainty. For both \Chandra\ energy bands, the power law slope is consistent with unity and the intrinsic scatter between \NuSTAR\ and \Chandra\ is negligible. However there is evidence for the \Chandra\ temperatures being systematically higher. We quantify this by using the best fitting model to calculate the percentage difference between the \NuSTAR\ and \Chandra\ temperatures at a \Chandra\ temperature of 10 keV.

We find that the broad-band \Chandra\ temperatures are on average $(10.5\pm3.7)\%$ higher at a \Chandra\ temperature of 10 keV than those measured with \NuSTAR\ (in the $3-10$ keV band). Similarly, the hard-band \Chandra\ temperatures are found to be $(15.7\pm4.6)\%$ higher than \NuSTAR\ temperatures at a \Chandra\ temperature of 10 keV.

\begin{figure}
    \includegraphics[width=\columnwidth]{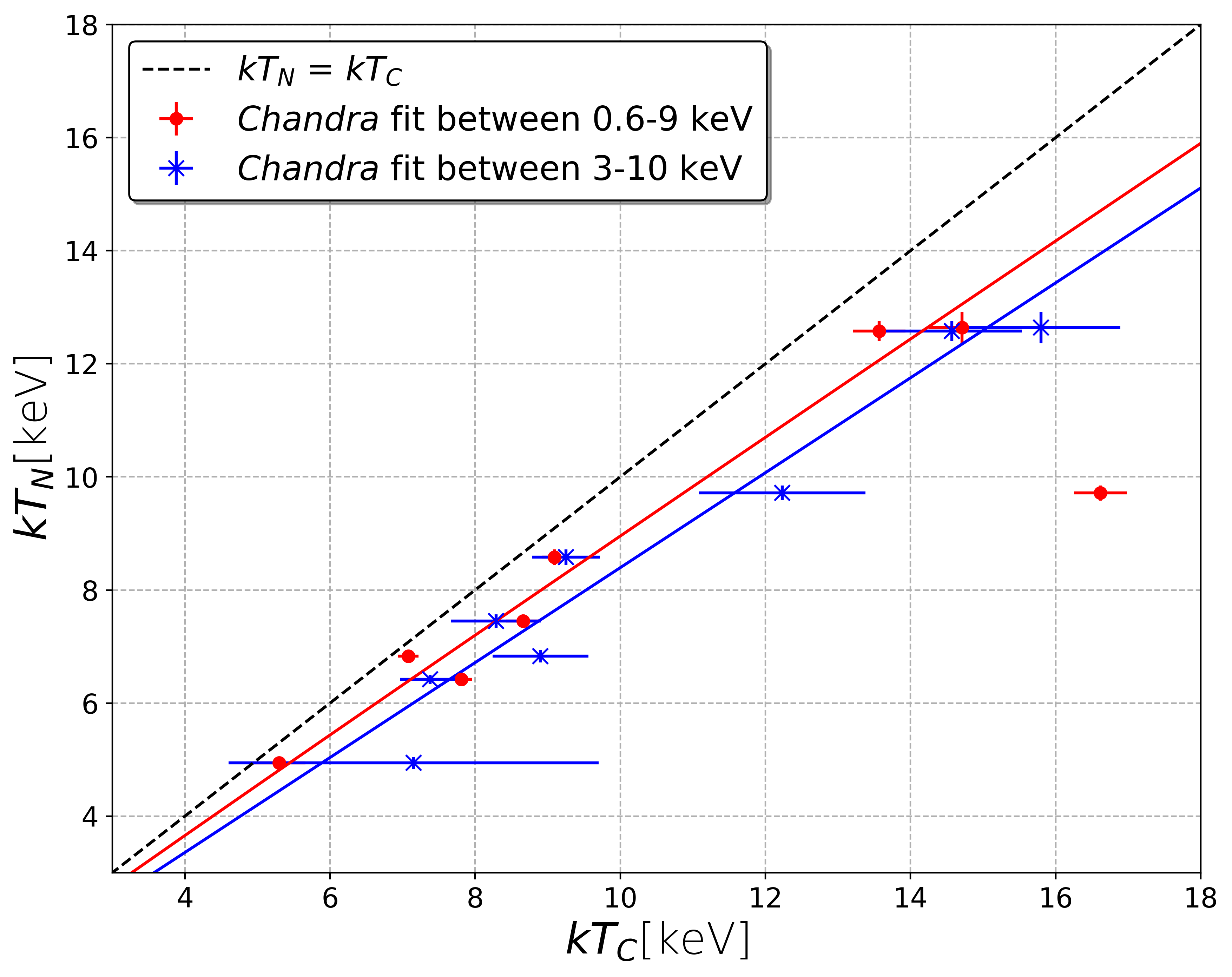}
    \caption{The cross-calibration between \textit{NuSTAR} and \textit{Chandra} for both broad-band (red points) and hard-band (blue points) \Chandra\ temperatures. The \textit{NuSTAR} spectra were fit between $3-10$ keV in all cases. The broad-band \Chandra\ temperature for Abell 2163 was excluded, as discussed in \textsection \ref{subsubsec:A2163}.}
    \label{fig:06_9v3_10_fit}
\end{figure}

Figure \ref{fig:log10_fit_v_XMM} also shows the cross-calibration model for \textit{Chandra} and \textit{XMM-Newton} temperatures derived by \citet{sch15} (for a different set of clusters). This gives an indication of the relative temperature calibrations of the three telescopes, and is discussed further in \textsection \ref{sec:disc}.
Figure \ref{fig:06_9v3_10_fit} shows the comparison between the cross-calibration fits for both the broad and hard-band \Chandra\ temperatures.

\section{Discussion}
\label{sec:disc}
Our comparison of cluster temperature measurements with \NuSTAR\ and \Chandra\ shows that the \Chandra\ temperatures are systematically higher. In the following we discuss possible sources of systematic uncertainty on our temperature measurements, and the limitations and implications of our cross-calibration measurement.

\subsection{Systematics on temperature measurements}
\label{subsec:syst}
We explored the possible impact on our results of uncertainties on the background modelling in the spectral analysis. We found that due to the high signal-to-noise of the cluster spectra, our results were insensitive to these systematics, but we briefly summarise here the main tests that were performed.

For the \NuSTAR\ analysis, the background spectrum is determined by fitting a model to spectra extracted from local background regions. In our standard analysis the background model is fit jointly to the spectra from all of the background regions, and the uncertainty on the resulting background model parameters is not propagated to the final cluster temperature. We tested the robustness of our temperatures by allowing the normalisations of the ``aperture'' background and ``fCXB'' components to fit independently to the spectra from each background region and then used the variation between regions as an estimate of the uncertainty on those components. The cluster temperatures were remeasured with the normalisation of those background components varied within these ranges.

For both the aperture and fCXB components, the resulting variation in cluster temperatures was less than or equal to the statistical uncertainty on the temperature measurement (with the exceptions of A523 where the aperture background gave a systematic of $\pm6\%$ on the temperature compared to a $2\%$ statistical error, and A2256 where the fCXB gave a systematic of $\pm4\%$ compared to a $1\%$ statistical error on the temperature. The change in normalisation of both background components is anti-correlated with the change in temperature, so if one of the background components were reduced by $1\sigma$ for all clusters, the temperatures would increase by an average amount that was less than the statistical errors. These systematics are thus subdominant to the systematic difference between \NuSTAR\ and \Chandra\ temperatures.

As discussed in \textsection \ref{subsec:nustar_data} we also considered the uncertainty on the \NuSTAR\ model background due to the possible presence of cluster emission in the background regions, and found this had negligible impact on the \NuSTAR\ temperatures. The inclusion (or not) of this extra background component produced a change in temperature that was always smaller than the statistical errors, and did not systematically increase or decrease the temperatures.

Furthermore, for all clusters, when the temperatures were measured independently for the FPMA and FPMB detectors, or independently for multiple observations, the agreement was good (e.g. the mean ratio of the FPMA to FPMB temperatures, computed in log space, was $0.99$ with a standard deviation of $0.06$).

For each cluster, the largest difference in temperature from any of the individual systematics investigated above was used as an estimate of the systematic temperature error for \NuSTAR. The cross-calibration fit was repeated using these systematic errors in place of the statistical errors on the temperatures. This made a negligible difference to the values of the best-fitting parameters or their uncertainties (as presented in table \ref{tab:fit}). We thus used only the statistical uncertainties on the temperatures in our analysis.

For \textit{Chandra}, the availability of blank-sky background spectra, and the ability to compare them with the in-field background meant the systematic uncertainties due to the background modelling were minimal.

\subsection{\Chandra/\NuSTAR\ cross-calibration}
\label{subsec:cross_cal_dis}
Our cross-calibration analysis shows that for all clusters in the sample, \NuSTAR\ measures their temperatures to be lower than \Chandra. This is unexpected, since the harder response of \NuSTAR\ should make it more sensitive to any hotter components present in the cluster spectrum. If both telescopes were perfectly calibrated then for any non-isothermal cluster, it would be expected that \NuSTAR\ would measure a higher temperature than \Chandra.

The use of different energy bands for the spectral fitting between the two telescopes would also contribute to systematic temperature differences. The broader band typically used for \Chandra\ spectral analysis ($0.6-9$ keV here), compared with the harder band ($3-10$ keV here) used for \NuSTAR\ should further exacerbate any temperature discrepancy, increasing the contribution of lower temperature components in a multiphase ICM to the global \Chandra\ temperature measurement.

Our results show that regardless of the energy band used, the \Chandra\ temperatures were systematically higher than those measured with \NuSTAR. When the broad band was used, the \Chandra\ temperatures were $(10.5\pm3.7)\%$ higher at a \Chandra\ temperature of 10 keV. When the hard band was used for the \Chandra\ analysis, the \Chandra\ temperatures increased (as would be expected) and were higher than the \NuSTAR\ temperatures by $(15.7\pm4.6)\%$ at a \Chandra\ temperature of 10 keV.

The fact that the temperature difference is significant when the data from both telescopes were fit in the hard ($3-10$ keV) band is interesting. This contrasts to some extent with previous cross-calibration work that has identified the calibration of X-ray telescopes at lower energies as largely responsible for temperature differences.

For example, \citet{sch15} found strong evidence that cluster temperatures from \Chandra\ and \XMM\ were inconsistent (at $>5\sigma$) when the spectra were fit in either soft ($0.7-2$ keV) or broad ($0.7-7$ keV) energy bands. When temperatures were measured in a hard ($2-7$ keV) band, the inconsistency was less significant (at the $\approx1-4\sigma$ level depending on the combination of \XMM\ detectors used). For these hard-band temperatures, the average \XMM\ temperature was between $0\%$ and $10\%$ lower than the average \Chandra\ temperature at a \Chandra\ temperature of $10$ keV.

The disagreement we find between \Chandra\ and \NuSTAR\ when \Chandra\ temperatures are measured in the hard band suggests that temperature difference is not driven by factors affecting the \Chandra\ measurements at low energies. This disfavours some possible origins for the temperature difference. For example, the uncertainty associated with the soft Galactic foreground modelling, or variations in the absorbing column (e.g. \textsection \ref{subsubsec:A2163} and \ref{subsubsec:A2256}) will not impact the temperatures measured in the hard band. Similarly, uncertainties on the modelling of the contamination build-up on \Chandra's optical blocking filter \citep{wei00} should not impact the hard-band temperatures.

\subsection{Implications for \Chandra/\XMM\ cross-calibration}
\label{sec:chandra_xmm}
Figure \ref{fig:log10_fit_v_XMM} presents the scaling relationship between \textit{Chandra} and \textit{XMM-Newton} derived in \citet{sch15} alongside the \textit{NuSTAR-Chandra} cross-calibration derived in our analysis. Here we show the ``ACIS-Combined XMM Full'' model from \citet{sch15}, measured in the $0.7-7$ keV energy band, while our \Chandra\ and \NuSTAR\ temperatures are measured in the $0.6-9$ keV and $3-10$ keV bands, respectively. The scaling relationship from \citet{sch15} is not directly comparable with our results due to a number of differences in the analyses (e.g. the comparison was made for different clusters, using a different definition for regions for the spectral extraction, and different energy bands for the spectral fitting), but it provides a useful indication of the relative direction and size of the the differences between the temperatures measured with different instruments.

Taken at face value, this comparison suggests that both \NuSTAR\ and
\textit{XMM-Newton} temperatures are systematically lower than those
measured with \Chandra\ (in the conventionally used broad band).
Furthermore, while we lack the data to directly compare \NuSTAR\ and
\XMM\ temperatures for the same clusters, the comparisons of each with
\Chandra\ in Figure \ref{fig:log10_fit_v_XMM} imply that the \NuSTAR\
temperatures would be systematically higher than \XMM. Hotter \NuSTAR\
temperatures are expected in the presence of multi-phase ICM, given
the harder \NuSTAR\ response relative to \XMM. The implication of this
indirect comparison (that \NuSTAR\ temperatures are hotter than \XMM\
but cooler than \Chandra) thus qualitatively favours consistency
between \NuSTAR\ and \XMM, with \Chandra\ being the outlier.

\begin{figure}
    \includegraphics[width=\columnwidth]{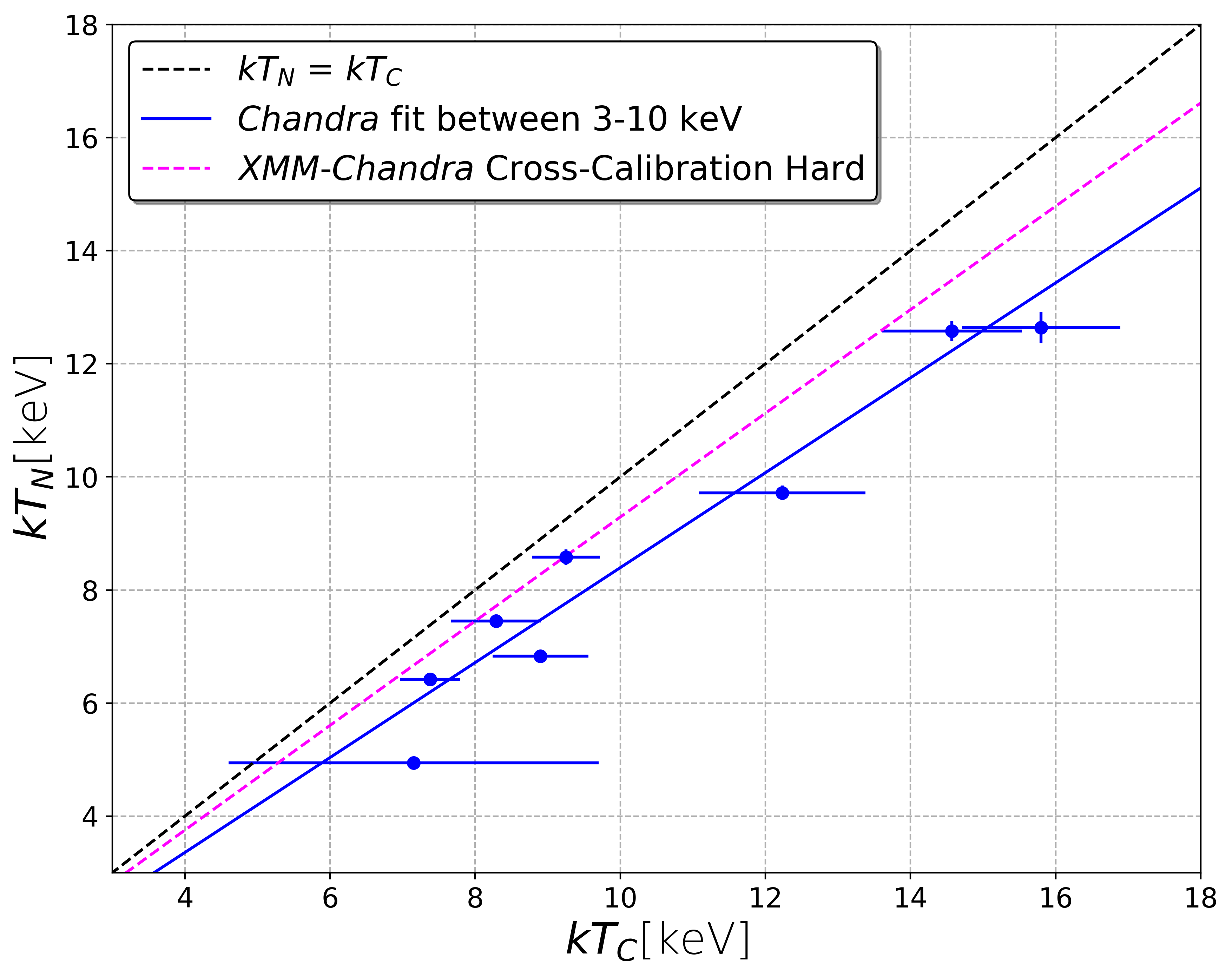}
    \caption{The cross-calibration between \textit{NuSTAR} and
        \Chandra\ temperatures measured in the $3-10$ keV band. The
        best fitting power-law model to the data is shown in blue, and
        the black dashed line indicates a perfect agreement between
        the temperatures. The dashed pink line shows, for illustrative
        purposes, the best fitting power-law to the cross-calibration
        of temperatures measured with \textit{XMM-Newton} versus
        \textit{Chandra} derived in \citet{sch15} (we show their
        ``ACIS-Combined XMM hard'' model, with \textit{XMM-Newton}
        temperatures in place of \NuSTAR\ temperatures on the vertical
        axis). Note that the comparison with the result from
        \citet{sch15} is not precise due to the different energy bands
        used for their temperature measurements, but illustrates the
        relative cross-calibration of the temperatures between the
        three telescopes.}
    \label{fig:log10_fit_v_XMM_hard}
\end{figure}

The picture is more complicated when hard-band temperatures are considered. This is illustrated in Figure \ref{fig:log10_fit_v_XMM_hard}, which shows the cross-calibration between \Chandra\ hard-band temperatures and \NuSTAR, along with the best-fitting relation between \XMM\ and \Chandra\ temperatures measured in the hard band from \citet[][we show their ``ACIS-Combined XMM Hard'' model, measured in the $2-7$ keV band]{sch15}. Again, this relation was derived for different clusters with a different hard-band definition, so is intended only as an illustration of the relative calibrations. This comparison implies that the hard-band temperatures measured with \XMM\ may be hotter than those measured with \NuSTAR, however the systematic differences between these studies make it difficult at present to explore this in more detail.

The relatively good agreement between hard-band \Chandra\ and \XMM\ temperatures found by \citet{sch15} and illustrated in Figure \ref{fig:log10_fit_v_XMM_hard} imply that \NuSTAR\ may be the outlier in this band. It is thus reasonable to consider whether the differences we find between \NuSTAR\ and \Chandra\ temperatures could be driven by uncertainties in the \NuSTAR\ calibration. However, in their recent \NuSTAR\ calibration update (CALDB 20211020), \citet{mad21} found good agreement between measurements of the Crab made with light that passed through the optics compared with stray light (which has a trivial effective area). Although we did not use CALDB 20211020 for our main analysis, we have verified that this CALDB update has negligible effect on our cluster temperatures, and so this increases confidence that the discrepancy we have found is not due to \NuSTAR\ calibration uncertainties.

Furthermore, two new (optional) updates to the \XMM\ effective area have recently been released which improve the internal agreement between the MOS and pn detectors, and also produce agreement with \NuSTAR\ in the hard band\footnote{\url{https://xmmweb.esac.esa.int/docs/documents/CAL-TN-0018.pdf}}. This is expected to move \XMM\ hard-band temperatures into better agreement with \NuSTAR. We thus conclude that the apparent agreement between hard-band \Chandra\ and \XMM\ temperatures, and discrepancy with \NuSTAR\ implied by the comparison between our work and \citet{sch15} does not constitute strong evidence that the difference is driven by uncertainties in the \NuSTAR\ calibration.

\subsection{Impact of multi-phase ICM}
\label{subsec:multi_phase}
As discussed previously, systematic differences in global temperatures are expected even for perfectly calibrated instruments if their energy responses differ and the source has multiple temperature components. It is therefore instructive to measure the temperature structure of the ICM in our target clusters. This was done by measuring radial temperature profiles from the \Chandra\ data.

The profiles were constructed by extracting and fitting spectra following the methods described in \textsection \ref{subsec:chan} and \ref{subsec:spec_fit}. Spectra were extracted from annular regions centred on the same location used for the global temperature measurement, with the widths of each annulus set such that the signal to noise of the resulting spectrum was at least 30.

The measured temperature profile of Abell 2256 is shown as an example in Figure \ref{fig:tp}, with the profiles of the other clusters shown in Appendix \ref{appen:A}. On the temperature profile plots we also show the global temperatures measured by \Chandra\ and \NuSTAR. For these comparisons the \Chandra\ temperatures were measured in the $0.6-9$ keV band, while the $3-10$ keV band was used for \NuSTAR.

\begin{figure}
    \includegraphics[width=\columnwidth]{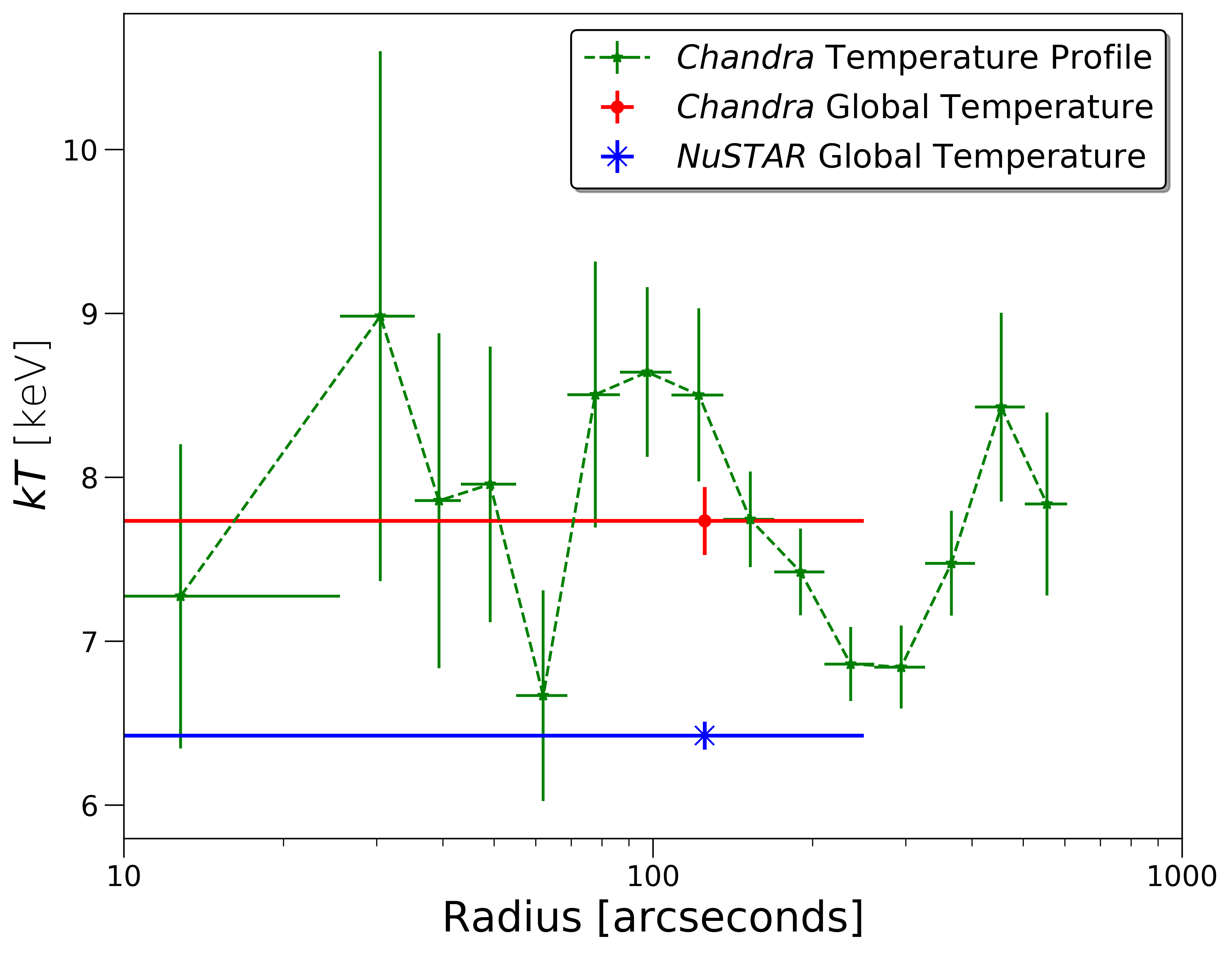}
    \caption{The \Chandra\ temperature profile of Abell 2256 is plotted in green. The global temperature across a circular region of radius 250 arcseconds is also shown for \textit{Chandra} (red) and \NuSTAR\ (blue). \textit{Chandra} and \NuSTAR\ temperatures are measured in the $0.6-9$ keV and $3-10$ keV bands respectively.}
    \label{fig:tp}
\end{figure}

The temperature profiles show that for each cluster, the global \NuSTAR\ temperature is lower than the \Chandra\ temperature found in the majority of the radial bins that overlap with the global temperature region. This underlines the systematic nature of the temperature difference. While each temperature bin will contain multiphase gas across its radial extent and projected along the line of sight, the profiles do not suggest that the global \Chandra\ temperature is higher than \NuSTAR\ due to their different weighting of different temperature components across the 250 arcsecond global temperature region.

In principle, the global \NuSTAR\ and \Chandra\ temperature could differ due to X-ray photons scattered into the global temperature extraction region by the large \NuSTAR\ PSF. In order for this to produce systematically lower \NuSTAR\ temperatures, there would need to be bright, cool regions of emission outside the 250 arcsecond global temperature region (possibly outside the \NuSTAR\ FOV). No such emission was apparent in the \Chandra\ images of these clusters, and the temperature profiles show no evidence for the systematic presence of large amounts of cool gas outside the global temperature extraction region.

It is clear from the temperature profiles of the clusters in our sample that they have complex thermal structures, which will contribute differently to the \NuSTAR\ and \Chandra\ global temperatures. The optimal calibration sources would be isothermal to avoid this effect. For example, \citet{sch15} removed the central regions of cool core clusters to create spectra that were closer to isothermality. For our sample, the clusters are mainly mergers, without strong cool cores, as can be seen in Appendix \ref{appen:A}, and so no core regions were excluded, and no attempt was made to select isothermal regions of the clusters.

We emphasise again that the multi-phase gas in these merging clusters should lead to \NuSTAR\ temperatures being higher than those from \Chandra\ due to the harder response of \NuSTAR. Thus, while the thermal structure of the ICM in our clusters means they are not ideal calibration sources, the systematic effect of this would be to lessen the disagreement we find between \Chandra\ and \NuSTAR. We can thus interpret the discrepancy we find as a lower limit on the possible disagreement between the temperature scale of the two instruments. A more definitive cross-calibration would require a sample of relaxed clusters for which isothermal regions can be defined. The required \NuSTAR\ observations are not currently available for a significant sample of relaxed clusters, and observing such targets would be of great value for the cross-calibration of current and future X-ray observatories.

\subsection{Implications of the \NuSTAR\ temperature calibration}
Our results imply a systematic uncertainty on X-ray temperature
measurements of clusters that is of order $10-15\%$. This is much
larger than the typical statistical uncertainty on temperatures for
clusters with good-quality data such as those studied here, and is one
of the dominant systematics for many applications of cluster
temperatures. For example, hydrostatic masses depend directly on
temperatures and their radial gradients so are directly impacted by
temperature systematics. \citet{sch15} showed that the difference
between their broad-band \Chandra\ and \XMM\ temperatures gave rise
hydrostatic masses that were $\approx15\%$ higher when assuming the
\Chandra\ calibration was correct compared to assuming the \XMM\
calibration was correct.

Measuring hydrostatic masses for our sample is beyond the scope of the
current paper (and the dynamically unrelaxed of the clusters make them
unsuitable for such analyses), but as discussed in \textsection
\ref{subsec:cross_cal_dis}, our \NuSTAR\ temperatures appear more
consistent with broad-band temperatures from \XMM\ rather than
\Chandra. We thus expect that assuming the \NuSTAR\ temperature
calibration would lead to results similar to those derived from \XMM\
temperatures by \citet{sch15}, with hydrostatic masses that were
$\sim15\%$ lower than those determined from \Chandra. This would then
mean that determinations of the hydrostatic bias derived from
\Chandra\ data would underestimate the level of hydrostatic bias by
$\sim15\%$. In other words, assuming the \NuSTAR\ temperatures were
correct would not reduce the amount of hydrostatic bias needed to
fully align the \textit{Planck} constraints on $\sigma_8$.

The impact of temperature calibration uncertainties was explored by
\citet{wan21} in the context of combining X-ray and Sunyaev-Zel'dovich
effect (SZE) measurements of the pressure in the ICM to constrain the
Hubble constant. \citet{wan21} found that the uncertainty on the
temperature calibration was the dominant systematic, with a $\sim10\%$
decrease in X-ray temperatures leading to an increase in $H_0$ of
$\sim10$ km s$^{-1}$ Mpc$^{-1}$. When using external priors on the
value of $H_0$, they found evidence that the \Chandra\ temperatures
used in their analysis were overestimated by $\sim10\%$ due to
calibration uncertainties. This is consistent with the results of our
comparison of \Chandra\ and \NuSTAR\ temperatures.

\section{Summary and Conclusions} \label{sec:summary}
We performed an X-ray spectral analysis on a sample of eight bright galaxy clusters to produce the first cross-calibration between \NuSTAR\ and \Chandra\ temperature measurements of the ICM. We found that \Chandra\ systematically finds ICM temperatures that are higher than those measured with \NuSTAR. We fit a power-law model to the temperatures and found that at a \Chandra\ temperature of 10 keV, the average \NuSTAR\ temperature was $(10.5\pm3.7)\%$ and $(15.7\pm4.6)\%$ lower than that of \Chandra, when the \Chandra\ spectra were fit in the broad ($0.6-9$ keV) and hard ($3-10$ keV) energy bands, respectively (the \NuSTAR\ spectra were always fit in the hard band).

We examined the impact of uncertainties on the background modelling and found that due to the high signal-to-noise of the cluster spectra, our results were insensitive to the details of the background modelling. We also examined the thermal structure of the clusters using temperature profiles from the \Chandra\ data, and found no evidence that \NuSTAR\ temperatures were being biased by emission from cool gas scattered from outside the spectral extraction region by the larger \NuSTAR\ PSF.

The fact that, when limited to the hard band for spectral fitting, the \Chandra\ temperature remains systematically higher than that from \NuSTAR, implies that the discrepancy is not driven by factors influencing the modelling at soft energies. These include systematics in the modelling of the absorbing column in the spectral analysis, and the calibration of the \Chandra\ effective area at soft energies to account for the ACIS contamination build-up.

We conclude that the difference is most likely due to systematic uncertainties in the calibration of one or both of the instruments. However, given that the presence of multiphase gas in the ICM is expected to lead \NuSTAR\ to measure a higher average temperature than \Chandra\ or \XMM, combined with previous findings that \Chandra\ temperatures are systematically higher than those of \XMM\ \citep{sch15}, we cautiously conclude that the evidence from \NuSTAR\ favours the \XMM\ temperature calibration when temperatures are fit in the conventional broad bands. However, the picture is less clear for temperatures measured in the hard band, and a more robust conclusion will require a direct comparison of \NuSTAR, \Chandra, and \XMM\ temperatures for the same clusters. Given that the sample used for the current analysis are thermally complex, merging systems, the definitive cross-calibration study should be deferred to a sample of relaxed clusters for which isothermal regions can be defined.

Our overall conclusion is that the average \NuSTAR\ temperature measurement is $10-15\%$ lower than that of \Chandra\ (at $10$ keV), most likely due to calibration differences between the observatories. Given the non-isothermal nature of the clusters studied, this is likely to be a lower limit on the difference in the relative temperature scales.

\section*{Acknowledgements}
BJM acknowledges support from STFC grant ST/R000700/1 and
ST/V000454/1. This research has made use of NASA’s Astrophysics Data
System Bibliographic Services and TOPCAT \citep{tay05}.

\section*{Data Availability}
All observations for the X-ray data analysis are publicly available at
the \textit{Chandra} and \NuSTAR\ Data Archives.

\bibliographystyle{mnras}
\bibliography{./bibly.bib}

\begin{appendices}
\section{Cluster images}
\label{appen:B}
\begin{figure*}
    \centering
    \subfloat{\includegraphics[scale=0.75]{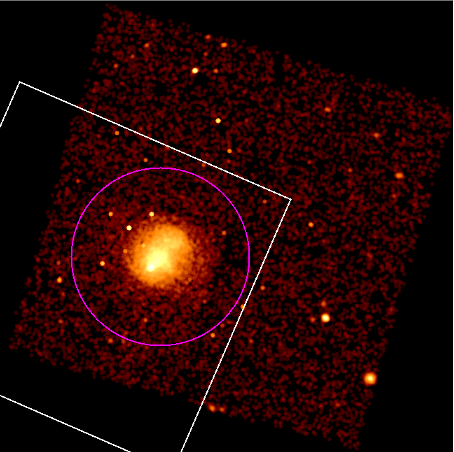}} \qquad
    \subfloat{\includegraphics[scale=0.75]{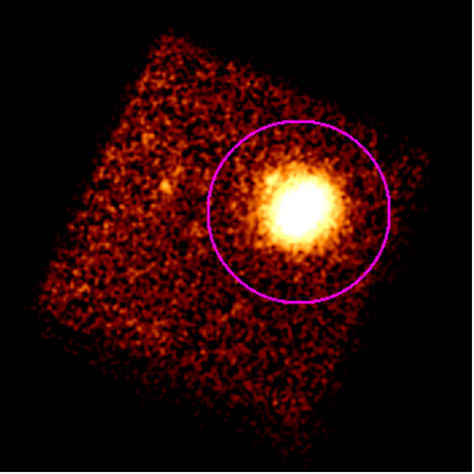}} \\ \textbf{A2146} \\
    \subfloat{\includegraphics[scale=0.75]{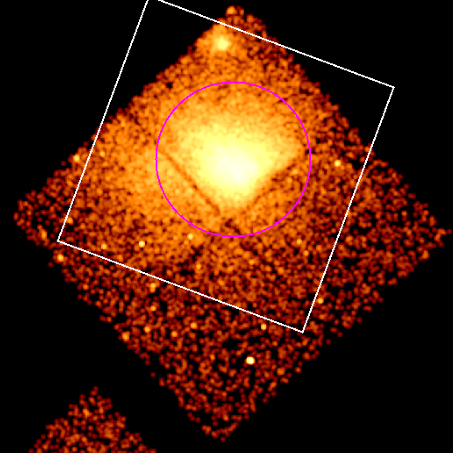}} \qquad
    \subfloat{\includegraphics[scale=0.75]{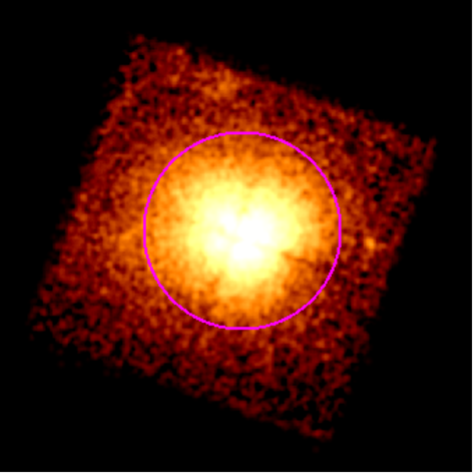}} \\ \textbf{A2163} \\     \subfloat{\includegraphics[scale=0.75]{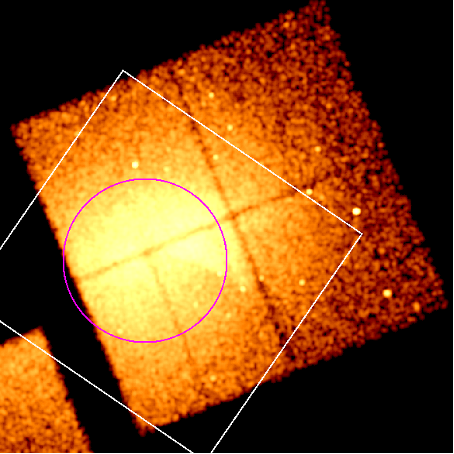}} \qquad
    \subfloat{\includegraphics[scale=0.75]{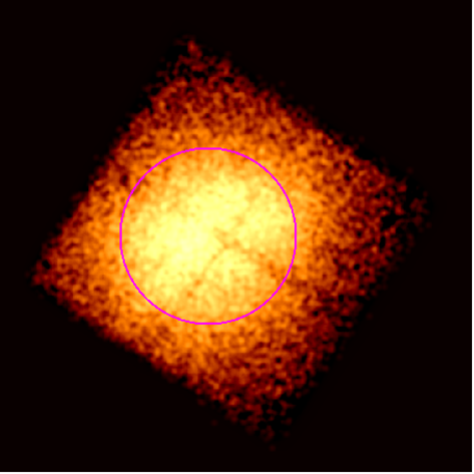}} \\ \textbf{A2256}  \\
     \caption{\Chandra\ (left) and \NuSTAR\ (right) images of A2146, A2163 and A2256. The \NuSTAR\ images are background subtracted and in the $3-20$ keV energy band, while the \Chandra\ image is in the $0.7-2$ keV energy band. All images are smoothed with a Gaussian with $\sigma=6.25"$. The magenta circle shows the region in which the source spectra were extracted. The white square shows the region covered by the \NuSTAR\ detectors overlaid on the \Chandra\ image. }
    \label{fig:app-images-1}
\end{figure*}

\begin{figure*}
    \centering
    \subfloat{\includegraphics[scale=0.75]{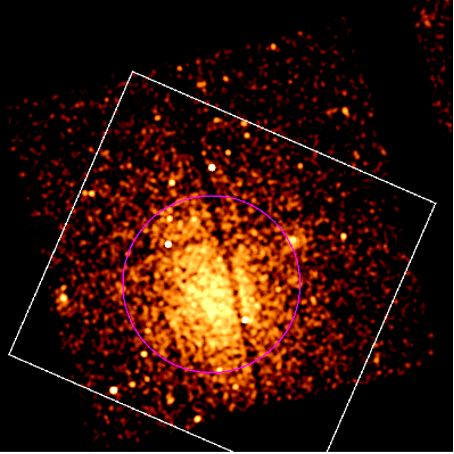}} \qquad
    \subfloat{\includegraphics[scale=0.75]{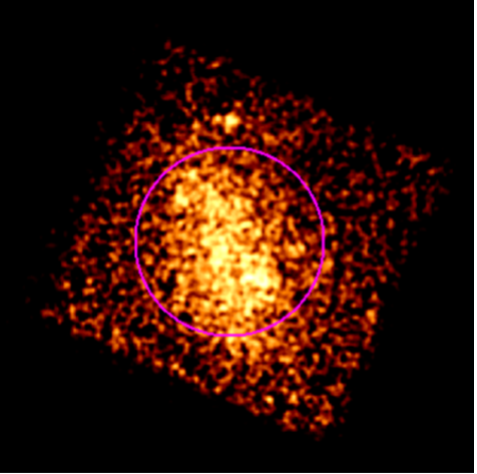}} \\ \textbf{A523} \\
    \subfloat{\includegraphics[scale=0.75]{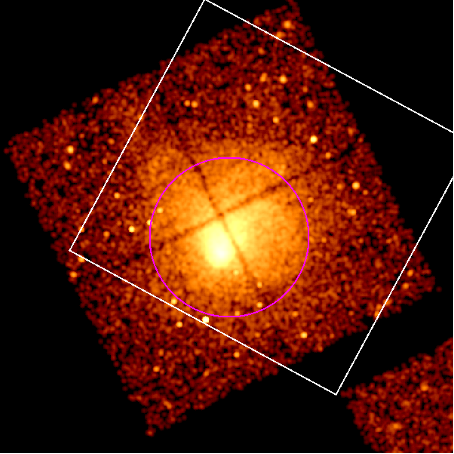}} \qquad
    \subfloat{\includegraphics[scale=0.75]{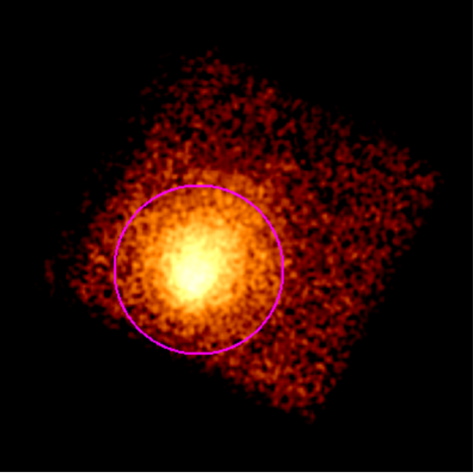}} \\ \textbf{A665} \\     \subfloat{\includegraphics[scale=0.75]{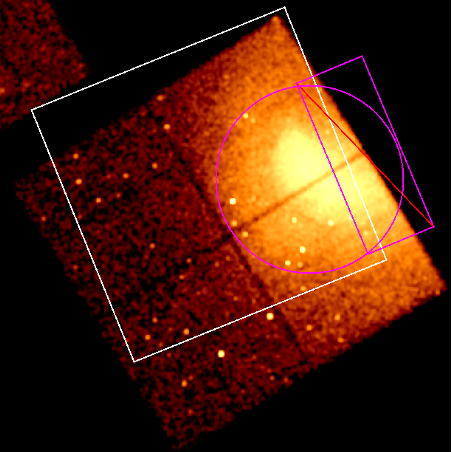}} \qquad
    \subfloat{\includegraphics[scale=0.75]{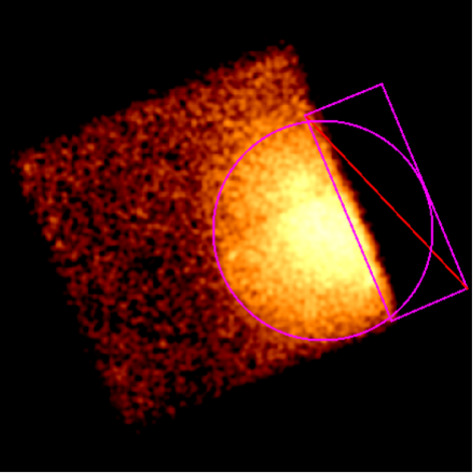}} \\ \textbf{A754}  \\
     \caption{\Chandra\ (left) and \NuSTAR\ (right) images of A523, A665 and A754. The \NuSTAR\ images are background subtracted and in the $3-20$ keV energy band, while the \Chandra\ image is in the $0.7-2$ keV energy band. All images are smoothed with a Gaussian with $\sigma=6.25"$. The magenta circle shows the region in which the source spectra were extracted. The white square shows the region covered by the \NuSTAR\ detectors overlaid on the \Chandra\ image.}
    \label{fig:app-images-2}
\end{figure*}

\begin{figure*}
    \centering
    \subfloat{\includegraphics[scale=0.75]{figs/bul_chan_250.png}} \qquad
    \subfloat{\includegraphics[scale=0.75]{figs/bul_6_250.png}} \\ \textbf{1E 0657-56} \\
    \subfloat{\includegraphics[scale=0.75]{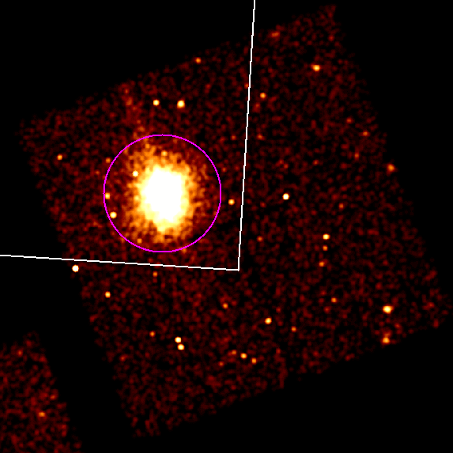}} \qquad
    \subfloat{\includegraphics[scale=0.75]{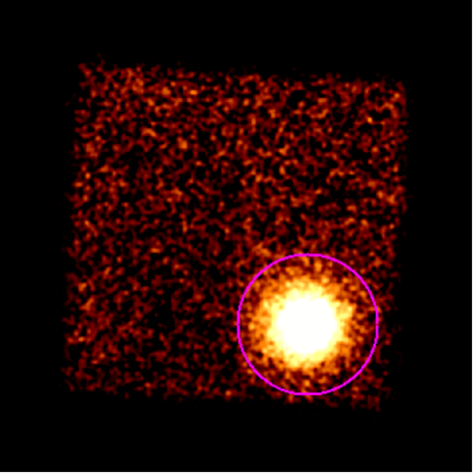}} \\ \textbf{RX J1347.5-1145}
     \caption{\Chandra\ (left) and \NuSTAR\ (right) images of 1E 0657-56 and RX J1347.5-1145. The \NuSTAR\ images are background subtracted and in the $3-20$ keV energy band, while the \Chandra\ image is in the $0.7-2$ keV energy band. All images are smoothed with a Gaussian with $\sigma=6.25"$. The magenta circle shows the region in which the source spectra were extracted. The white square shows the region covered by the \NuSTAR\ detectors overlaid on the \Chandra\ image.}
    \label{fig:app-images-3}
\end{figure*}

The \Chandra\ and \NuSTAR\ images of the eight clusters in the sample are shown in Figures \ref{fig:app-images-1} - \ref{fig:app-images-3}. In each image, the magenta circle shows the region in which the source spectra were extracted. The white square shows the region covered by the \NuSTAR\ detectors overlaid on the \Chandra\ image.


\section{Temperature profiles}
\label{appen:A}
\begin{figure*}
    \centering
     \subfloat{\includegraphics[width=\columnwidth]{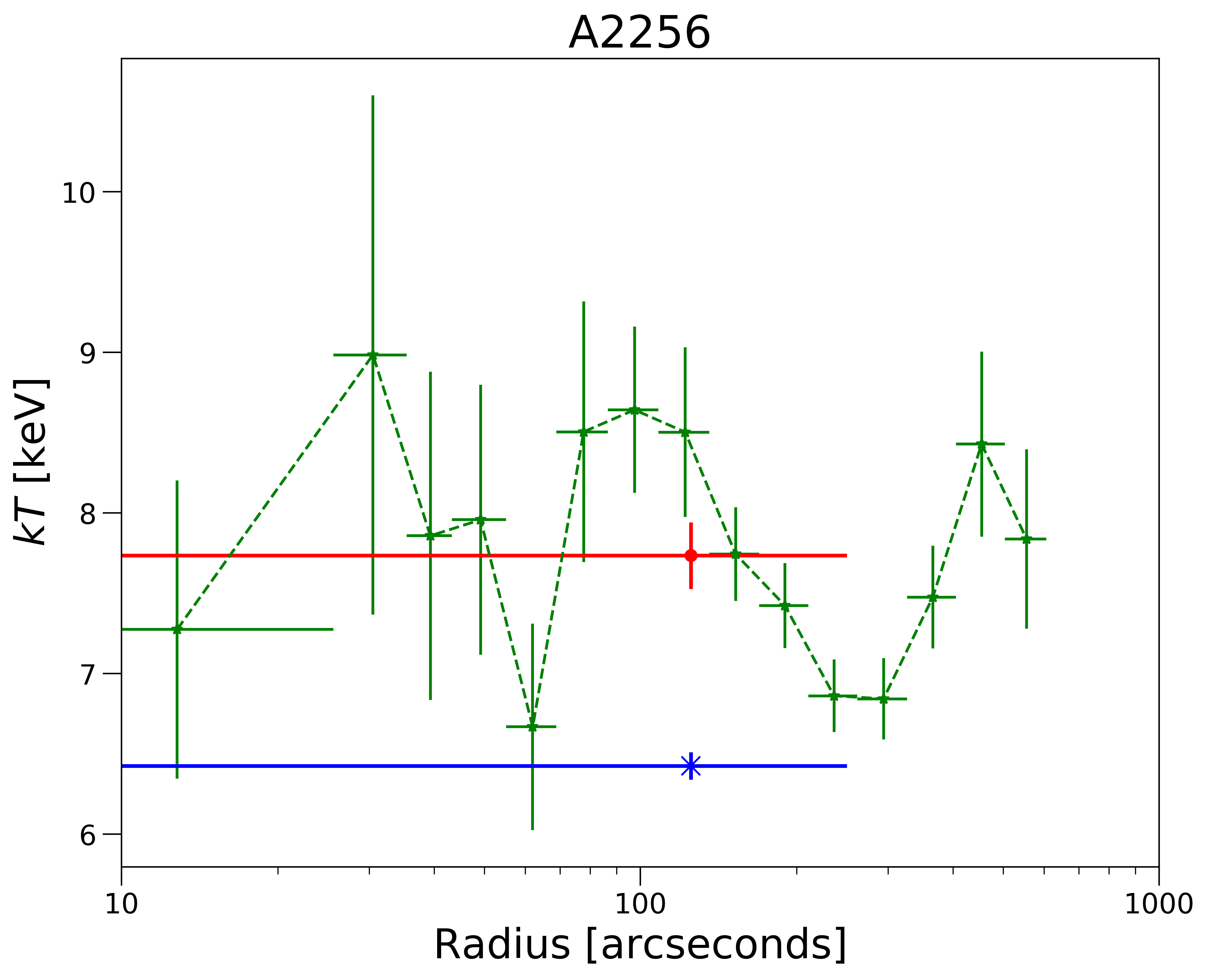}}
     \subfloat{\includegraphics[width=\columnwidth]{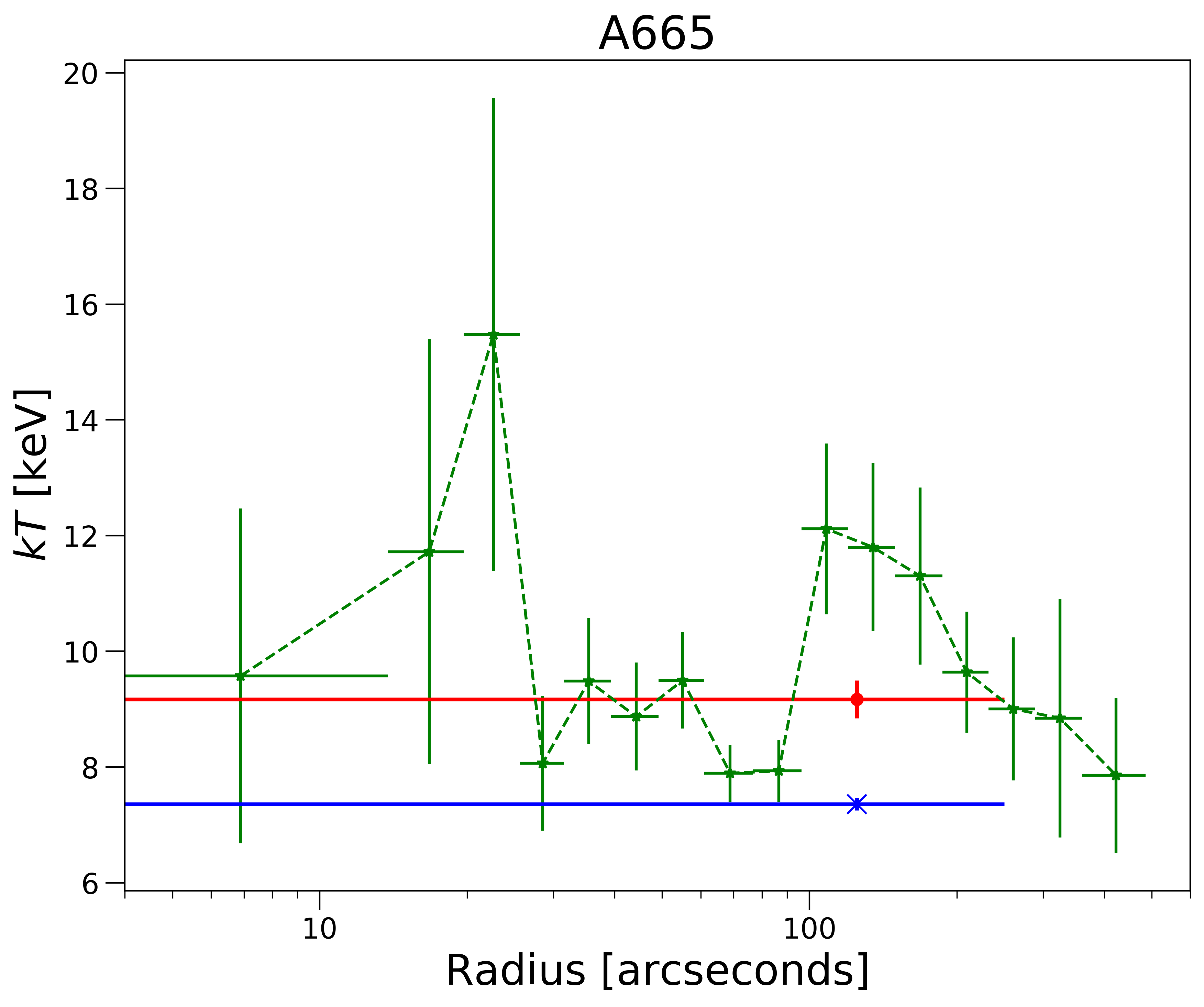}}\\
     \subfloat{\includegraphics[width=\columnwidth]{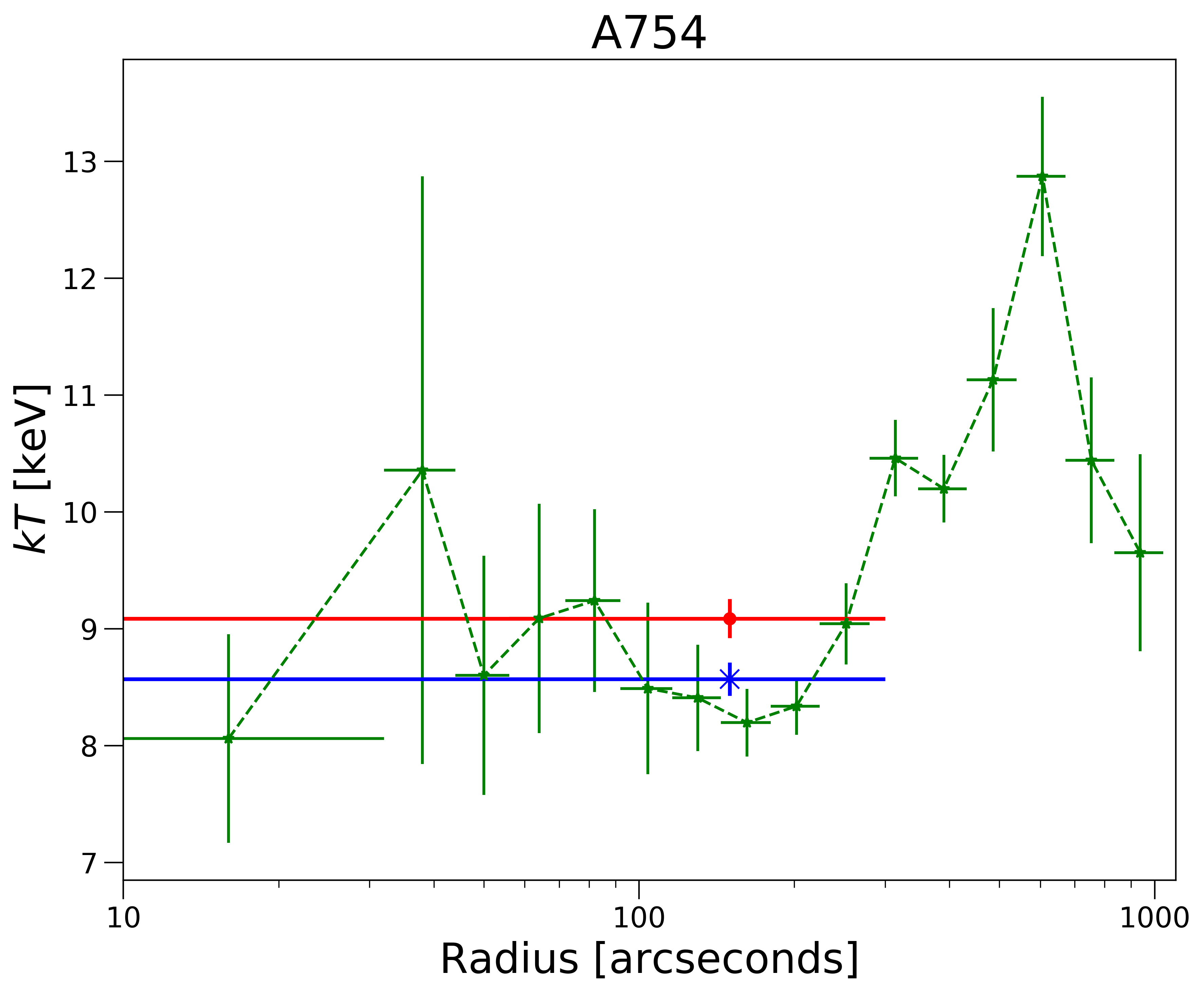}}
     \subfloat{\includegraphics[width=\columnwidth]{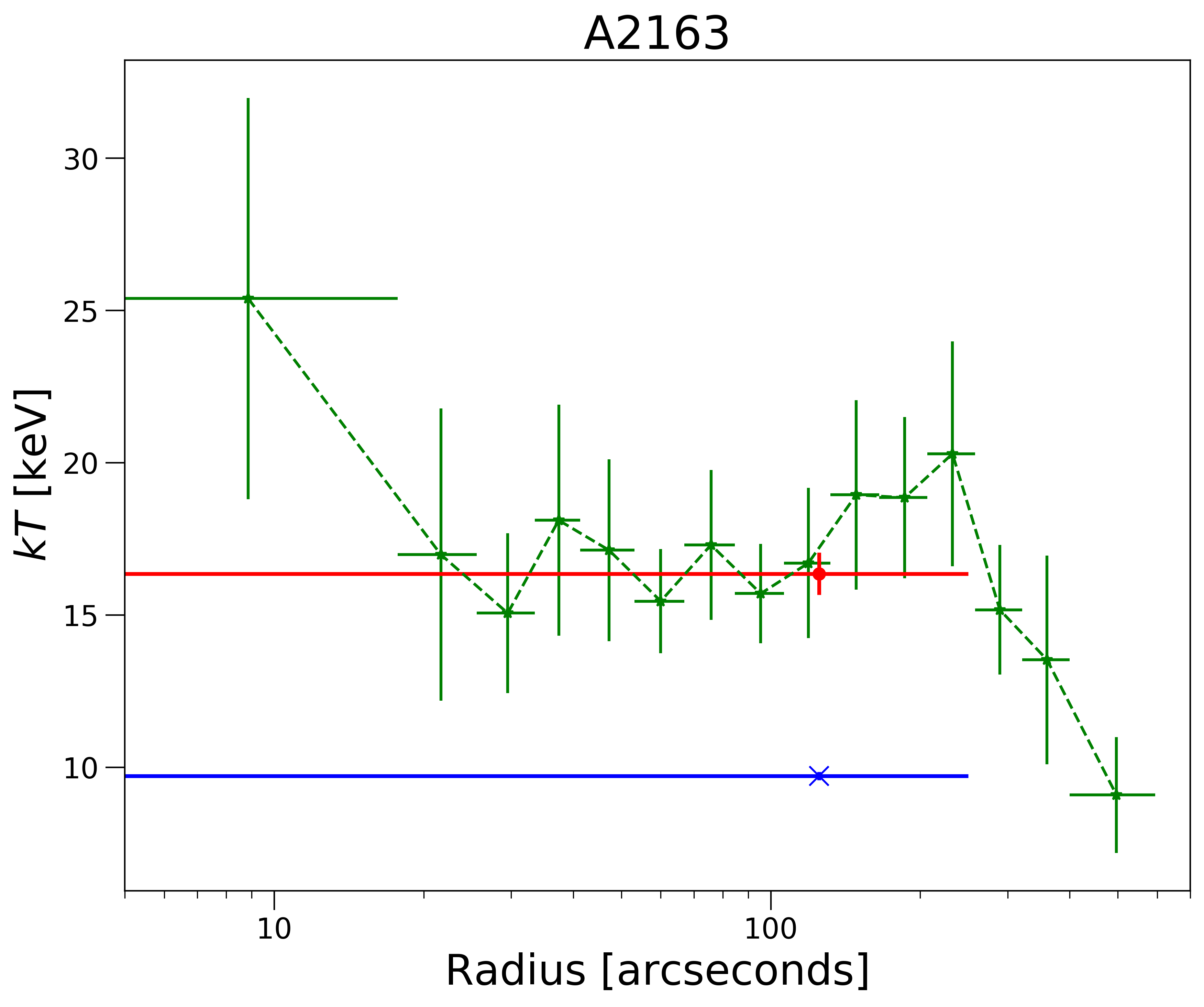}}
    \caption{Temperature profiles of clusters A2256, A665, A754 and A2163 plotted in green. The global temperature across a circular region of 250 arcseconds is shown for \textit{Chandra} (red) and \NuSTAR\ (blue). \textit{Chandra} and \NuSTAR\ temperatures are measured in the $0.6-9$ keV and $3-10$ keV bands respectively. Notes that due to problems in the background of the Abell 2163 \textit{Chandra} observation at low energies (\textsection \ref{subsubsec:A2163}), the global temperature from the fit between 0.6-9 keV was not used in our main analysis. The temperature fit between 3-10 keV (pink) was alternatively plot.}
    \label{fig:rp_1}
\end{figure*}

\begin{figure*}
    \centering
     \subfloat{\includegraphics[width=\columnwidth]{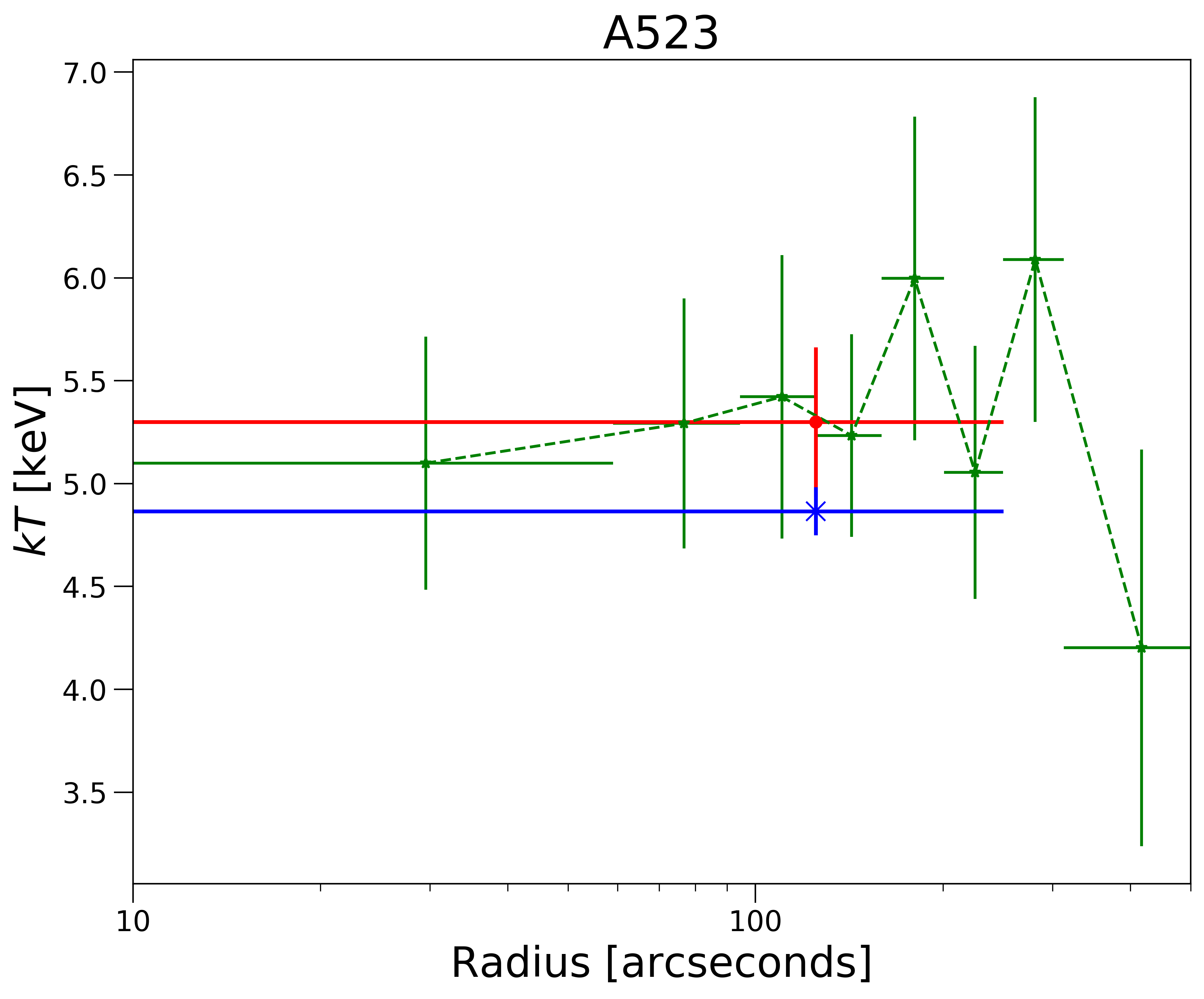}}
     \subfloat{\includegraphics[width=\columnwidth]{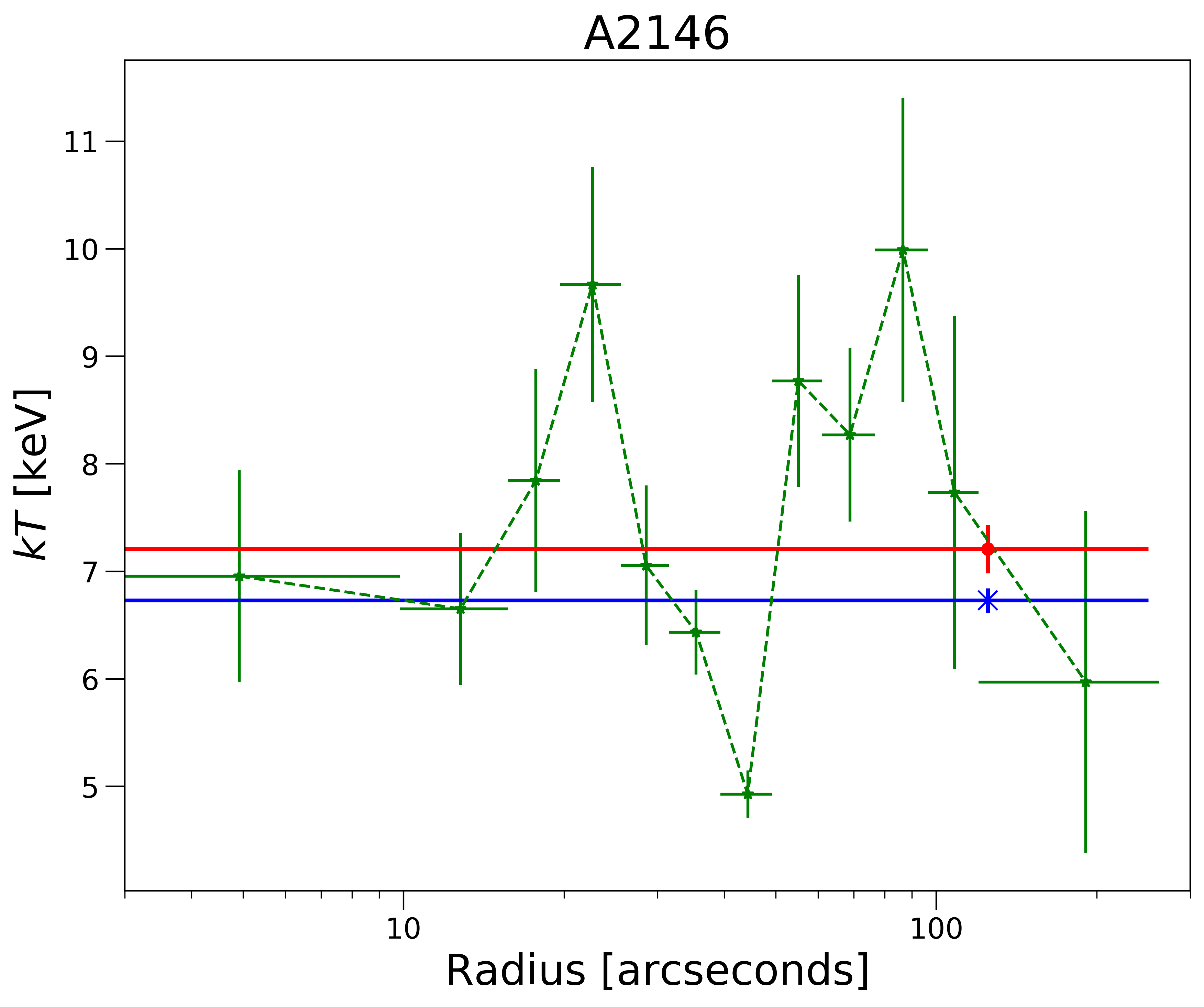}} \\
     \subfloat{\includegraphics[width=\columnwidth]{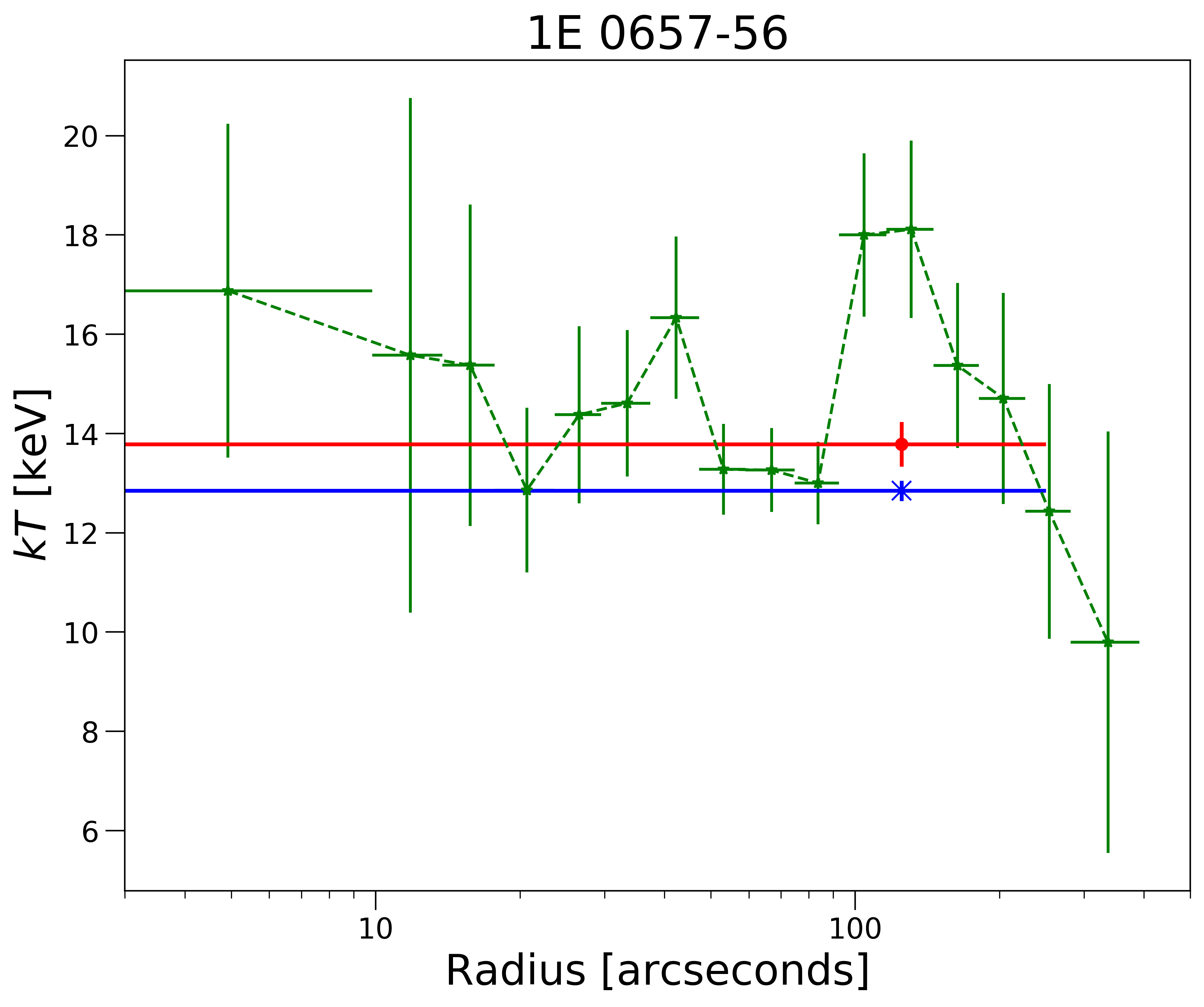}}
     \subfloat{\includegraphics[width=\columnwidth]{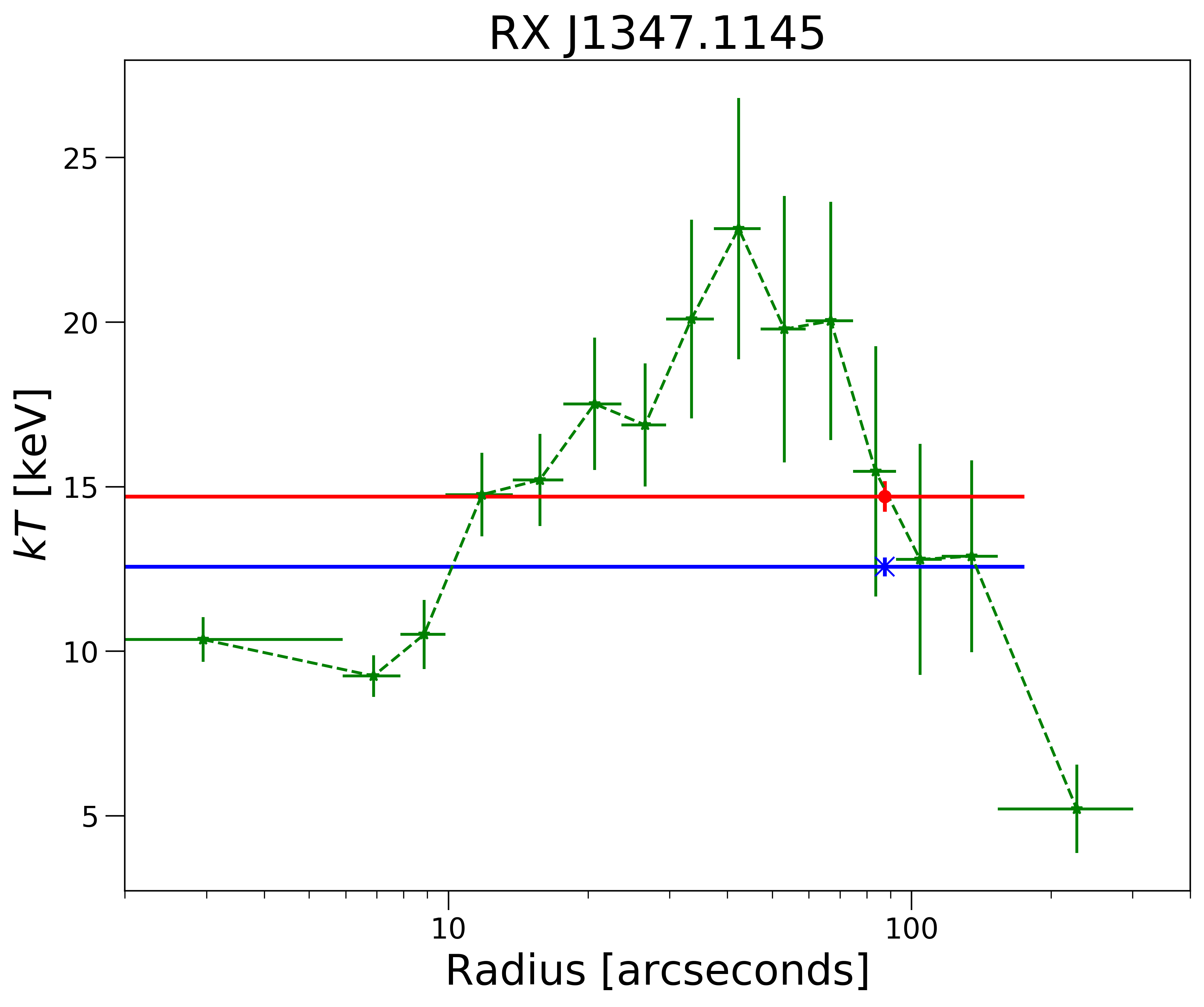}}
    \caption{Temperature profiles of clusters A523, A2146, 1E 0657-56 and RX J1347.5-1145 plotted in green. The global temperature across a circular region of 250 arcseconds is shown for \textit{Chandra} (red) and \NuSTAR\ (blue). \textit{Chandra} and \NuSTAR\ temperatures are measured in the $0.6-9$ keV and $3-10$ keV bands respectively.}
    \label{fig:rp_2}
\end{figure*}

Figures \ref{fig:rp_1} and \ref{fig:rp_2} show the temperature profiles created using \Chandra\ for the eight galaxy cluster in this sample, following the methods described in \textsection \ref{subsec:chan}, \ref{subsec:spec_fit} and  \ref{subsec:multi_phase}. Overlayed are the global temperatures of the clusters measured by \Chandra\ and \NuSTAR\ across the chosen source region.

\end{appendices}

\bsp	%
\label{lastpage}
\end{document}